\documentclass[twocolumn]{aa}
\usepackage{natbib}
\bibpunct{(}{)}{;}{a}{}{,}
\usepackage{psfig}
\usepackage{epsfig,afterpage} 
\usepackage{graphicx}
\usepackage{pifont}
\usepackage{amssymb}
\usepackage{longtable}
\usepackage{array}
\usepackage{amsmath}
\usepackage{rotating}
\usepackage{multirow}
\usepackage{lscape}

\begin{document}
\title{MULTIVARIATE ANALYSIS OF GLOBULAR CLUSTER'S HORIZONTAL BRANCH MORPHOLOGY: searching
for the second parameter.
\thanks{Based on observations with the {\it Hubble Space telescope + WFPC2}}} 

\author{A. Recio-Blanco\inst{1,3},  A. Aparicio\inst{2,6}, G. Piotto\inst{3}, F. De Angeli\inst{3,5},
S.G. Djorgovski\inst{4}}

\offprints{A. Recio-Blanco}

\institute {Dpt. Cassiop\'ee, UMR 6202, Observatoire de la Cote d'Azur,
B.P. 4229, 06304 Nice Cedex 04, France\\
\email{arecio@obs-nice.fr}
\and 
Departamento de Astrof\' isica, Universidad de La Laguna, V\' ia L\'actea s/n, 38200, 
La Laguna, Tenerife, Spain\\
\email{antapaj@iac.es}
\and 
Dipartimento di Astronomia, Universit\`a di  Padova,   
Vicolo  dell'Osservatorio  2, I-35122 Padova, Italy\\
\email{recio,piotto,deangeli@pd.astro.it }
\and
Astronomy Department, California Institute of Technology, MC 105-24, Pasadena, CA 91125.\\
\email{george@astro.caltech.edu}
\and
Institute of Astronomy, Madingley Rd, CB3 0HA Cambridge, UK\\
\email{fda@ast.cam.ac.uk}
\and
Instituto de Astrof\' isica de Canarias, V\' ia L\'actea s/n,
38200 La Laguna, Tenerife, Spain\\
}

\date{Received 8th March 2005; accepted 16th November 2005}

\abstract{}{The interpretation of globular cluster horizontal branch (HB) morphology is
a classical problem that can significantly blur our understanding of stellar populations.}
{In this paper, we present a new multivariate analysis connecting the effective temperature
extent of the HB with other cluster parameters. The work is based on Hubble Space Telescope 
photometry of 54 Galactic globular clusters.}{
The present study reveals an important role of the total mass of the globular cluster 
on its HB morphology. More massive clusters tend to have
HBs more extended to higher temperatures.  For a
set of three input variables including the temperature extension of the HB, $[Fe/H]$
and M$_V$,  the first  two eigenvectors account  for  the 90\%  of the total  
sample  variance.}{  Possible effects of cluster self-pollution on HB morphology,
eventually stronger in more massive clusters, could explain 
the results here derived.}

\keywords{globular clusters: general --- stars: horizontal-branch --- stars: Population II}

\authorrunning{Recio-Blanco et al.\ }

\titlerunning{HB Multivariate analysis }

\maketitle

\section{Introduction}

Globular  clusters (GC), comprised  of chemically  homogeneous  and coeval
populations of stars, represent excellent systems for testing
stellar   models.  The various sequences    that appear in the
colour-magnitude  diagram  (CMD) of a globular cluster can  be  compared  to  the predicted
isocrones and  theoretical loci. In this way,  the properties of stars
at  different stages of  evolution,  and the fundamental characteristics of
the clusters  themselves,  such as cluster distance   and  age, can be
derived. Hence, it is not surprising that   the study of GCs has played a
pivotal role in the development of stellar astrophysics.

In this paper, we will focus  on one evolutionary
stage: the  horizontal  branch. HB stars  are  characterized by
core-helium burning and  shell-hydrogen burning.  The star location in
effective temperature along   the  Zero Age Horizontal Branch   (ZAHB)
depends   on  virtually  all   stellar  parameters  (composition, age,
rotation, etc..., see e.g.  Rood 1973).  The wide  colour distribution
of  the HB,  called  the HB morphology,  is the result of  large
differences in the  envelope mass of  stars having the same core mass,
at  the same  evolutionary stage. The  HB phase behaves  as an amplifier,
displaying the record of both initial conditions and of any variations
and perturbations  in the evolution of the  star from its  birth up to
the HB stage. Therefore, reading properly the  HB morphologies can yield a
better understanding of   Population II stellar  evolution in general,
and of the specific stellar systems, clusters or galaxies, in particular.

However, it  appears that our comprehension  of  the HB  phase and its
precursors is incomplete.  Canonical stellar theory can not adequately
explain the wide variety of  HB morphologies observed in Galactic GCs,
ranging from short red HBs to  long extended ``blue  tails''.  In
particular,  blue tails still  represent a puzzle   in the stellar  evolution
model, in the sense that we know what the stars in  blue tails are,
but  we do not  know how stellar evolution can  create them. 

To a first approximation,   the  different temperature extension   and
morphology  of  the  observed  HBs  were  interpreted  in terms  of  the
metal abundance variation, the  {\it first  parameter}: metal-rich
clusters tend  to have  short red HBs,  while metal-poor  ones exhibit
predominantly blue HBs. Nevertheless, the previous approximation turned out
to  be  too  rough   very  soon.  Some   other   parameter (or set  of
parameters) was evidently    also at work,  as    clusters with nearly
identical  metallicities    could  show  very  different   HB   colour
distributions (van den Berg  1967, and Sandage \&  Wildey 1967). One
classical example is the  pair formed  by M3 and  M13, with $[Fe/H]=-1.57$ 
and $[Fe/H]=-1.54$ (Harris 1996, in its revised version of 2003)
respectively, but very different HB morphologies.
The variety of proposed candidates range from cluster age, to helium mixing, [CNO/Fe] 
abundance, cluster concentration, stellar rotation, planets...
In fact, the second parameter problem has already been the object of an extensive list 
of studies that is important to acknowledge. An increasing amount of 
observational data progressively revealed the complexity of the scenario (Kraft 1979,
Freeman \& Norris 1981). Important work was developed by the Yale group
(e.g. Lee et al. 1987, 1988, 1990, Sarajedini \& King 1989) interpreting
cluster age as a {\it global} second parameter,
in the framework of a Galactic formation picture a la Searle \& Zinn (1978).
On the other hand, relevant questions outside this scenario were also
discussed by different authors (Renzini \& Fusi Pecci 1988, Rood \& Crocker 
1989, Buonanno et al. 1989, Fusi Pecci et al. 1990,  Fusi Pecci et al. 1993).
In particular, the idea that production of hot HB stars may be
somehow influenced by the dynamical processes in the cluster was also carefully 
explored. In agreement with this, Fusi Pecci et al. (1993) and Buonanno et al. 
(1997) suggested environement density as a possible second paramenter. Finally, among other
more recent works those of Soker \& Harpaz (2000) and Catelan et al. (2001)
can be cited.

In this work, we have analyzed a homogeneous database of 54 globular cluster CMDs
from Hubble Space Telescope photometry in an attempt to quantify the different
dependences of HB morphology on cluster parameters: metallicity,
concentration,  distance to the Galactic  center, total mass, etc...,  in
other  words, to search for the  so  called HB second parameter(s).
The data, reduced and treated in  an  uniform way, represent  an  exceptional
opportunity, from the statistical point of view, to investigate how HB
morphology   depends  on globular   cluster   properties. On the other hand,
the multidimensional data set of Galactic globular  clusters spans a 
large  range in many  of their properties such as luminosity, metallicity, 
etc... Therefore, in  order to  reveal the possible  complex correlations  
hidden  in  the  HB second parameter problem,  we decided to apply a  
multivariate statistical analysis. This approach can not only confirm
or reveal new correlations, but offers the possibility to estimate the
relative importance of the various HB dependences and the degree of explanation 
of HB morphology that can be obtained by their combination.
A similarly motivated study was already done by Fusi Pecci et al. (1993).
We believe that our new, HST-based data set warrants a fresh
look at the problem.

\section{The database}

The analysis presented  here is based on a \textit{Hubble
Space  Telescope} snapshot program aimed  at mapping  the cores of all
GCs with (m-M)$_B$ $<$ 18.0, using the Wide  Field/ Planetary Camera 2
(WFPC2) of  HST.   All    the   photometric data come  from HST/WFPC2
observations  in the $F439W$ and $F555W$   bands, the WFPC2 equivalents of
the $B$   and  $V$ filters,  which  are suited  for a  generic  survey and
constitute the  best choice  to identify new  anomalous HBs.   In  all
cases, the  PC camera was centered  on the  cluster center.

The  colour-magnitude   diagrams   and    the
photometry derived  from   that program have been already published by
Piotto et al.\ (2002). Moreover, the  database has already given rise
to  a number of works attacking  still-open topics on evolved stars in
GCs (Piotto  et al.\ 1999,   Palmieri et al.\ 2002, Raimondo et al.\  2002, Zoccali  et al.\
1999, Bono et al.\  2001, Zoccali et al.\  2000, Cassisi et al.\ 2001,
Riello et al. 2003, Recio-Blanco et. al. 2004, Piotto et al. 2004,
Salaris et al.\ 2004, De Angeli et al.\ 2005 ). The complete database
consists in a total of 74 GCs (53 snapshot plus 21 archive data). For
this work, only those clusters whose CMD had a well populated HB
and enough photometric precision to offer reliable estimations of HB 
temperatures, 54 of them, were used. 

Due to the  severe stellar crowding,  the excellent resolving power of
HST  is crucial for the photometrical  studies of the globular cluster
central  regions,    where accurate  ground  based    observations are
precluded.   Moreover,  for GCs near   the  Galactic center, only  the
central regions  are  sufficiently uncontaminated  by  field stars  to
allow a   good study of the   colour-magnitude  diagram.

\section{Considered cluster parameters}

\subsection{The morphology parameter: maximum effective temperature along
the HB}

The  first step  in  this  work consists   in  the evaluation of  the HB 
morphology of each cluster. In order  to have a quantitative  measure of
their extension,   we  determined the  highest  effective  temperature
reached by the stars in the HBs of all the clusters in our data set by
fitting a   Zero Age Horizontal Branch (ZAHB)    model to the observed
CMDs.  In this way,  we can study how the  extension of the horizontal
branch  varies with cluster  parameters.   It is  worth to mentioning,
however, that HB bimodality, i.e. the presence of  both a red HB and a
blue tail  as in  NGC2808,  is not been  taken  into account with this
approach.

 ZAHB models from Cassisi et al. (1999) were fitted to our $F439W,F555W$ 
CMDs using the values  of  distance modulus and reddening derived in our 
previous paper, Recio-Blanco et  al.\ 2004, for each of  the clusters in  
the data set.

\begin{figure*}
\centerline{\includegraphics[width= 13 cm]{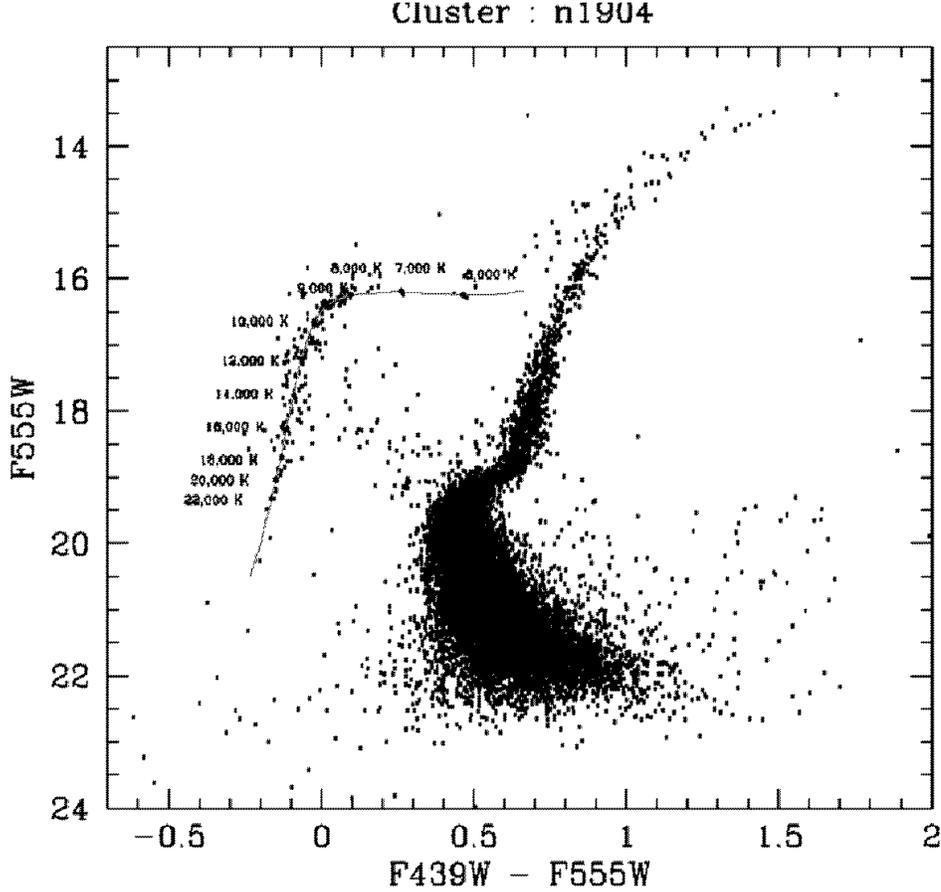}}
\protect\caption[]
{ZAHB model by Cassisi et al. (1999) fitted to NGC 1904
colour-magnitude diagram. The effective temperature variation
along the HB is also shown.}
\label{firstp}
\end{figure*}

This procedure allowed  us to evaluate the highest $T_{\rm eff}$
reached   by the globular cluster   HB  and therefore its  temperature
extension,  as   illustrated in    Figure 1 for the case of NGC~1904
for which the corresponding temperatures along the HB are marked. 
The  errors in  this
temperature determination are difficult to estimate as they depend not
only on the errors  in the distance modulus, $(m-M)_{F555W}$, and  reddening, but also on the
number of stars in the  HB and the temperature  range we have to  deal
with. As a consequence, the largest errors occur for the smallest low
central  concentrated clusters, and for  the most  extended HBs, where
the  large bolometric correction in  these photometric bands precludes
an accurate estimation of $T_{\rm eff}$.  However, although the errors can
be  rather important, the general  trend of HB morphology with cluster
parameters is not   dramatically affected, as we  will  see in 
the next Section.  

Finally, we have performed a comparison between our $T_{\rm eff}$
parameter, in its logarithmic form, and the L$_t$ parameter
from Fusi Pecci et al. (1993). As explained in their Section 3.3.1,
the L$_t$ parameter measures the total lenght of the HB from the
HB red endpoint. The result of the comparison is presented in Figure 2.
As indicated, the derived correlation is 0.65. In fact, although there is
a clear common trend, the spread of the points is rather high, most
probably due to the difference in the available photometric sources. 
Fusi Pecci et al. (1993) measurements come from an extensive, although
not homogeneous, collection of ground based CMDs from 1965 to 1992. 
On the contrary, our measurements were performed using recent homogeneous HST data base.
Therefore, alghouth both parameters rely on a somewhat subjective
estimation of the terminal HB point, an important part of the observed
rms is probably coming from the photometric data.

\begin{figure*}
\centerline{\includegraphics[width= 13 cm]{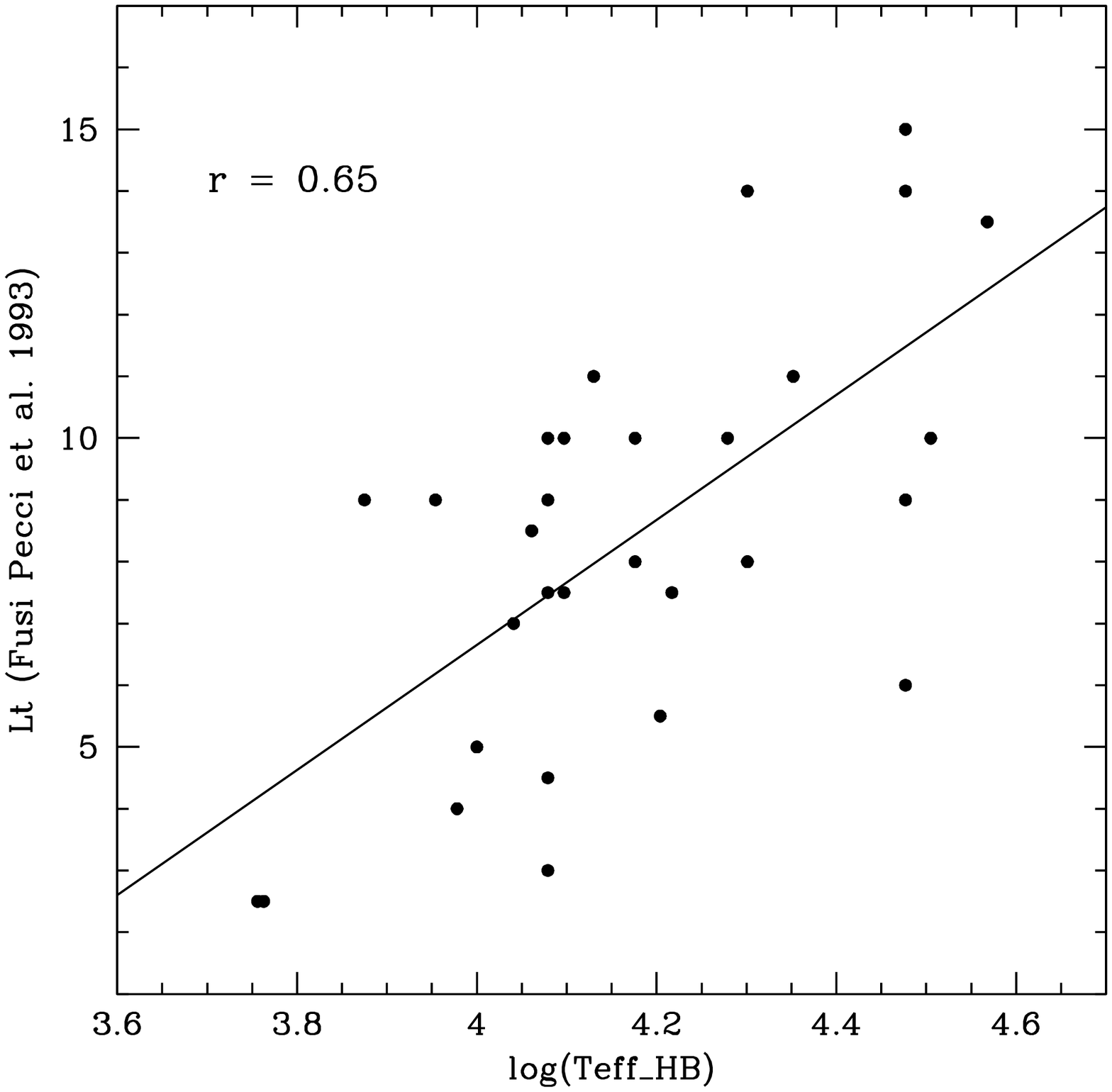}}
\protect\caption[]
{Comparison between our $T_{\rm eff}$
parameter, in its logarithmic form, and the L$_t$ parameter
from Fusi Pecci et al. (1993).}
\label{firstp}
\end{figure*}

%\begin{sidewaystable}
\begin{table*}
\centering
%\begin{sideways}
\scriptsize{\begin{tabular}[b]{cccccccccccccccc} 
\hline
\hline  
Id&\tiny{log(Teff$_{HB}$)}& $[Fe/H]$& M$_V$ &$\Gamma_{col}$&$\rho_{0}$&c&R$_{GC}$&    l    &    b    & r$_{c}$&r$_{h}$& t$_{rc}$ &t$_{rh}$& $\mu_{V}$ & Age$_{rel}$\\      
\hline
 NGC0104  &   3.756  & -0.76  & -9.42  & -13.64 & 4.77 & 2.03  &  7.4 & 305.90 & -44.89   &  0.44 & 2.79 &   8.06  &  9.48 &  14.43 & 0.97\\
 NGC0362  &   4.079  & -1.16  & -8.40  & -13.58 & 4.70 & 1.94  &  9.3 & 301.53 & -46.25   &  0.19 & 0.81 &   7.76  &  8.92 &  14.88 &  0.74\\
 NGC1261  &   4.079  & -1.35  & -7.81  & -14.70 & 2.96 & 1.27  & 18.2 & 270.54 & -52.13   &  0.39 & 0.75 &   8.74  &  9.20 &  17.65 & 0.75 \\
 NGC1851  &   4.097  & -1.22  & -8.33  & -13.28 & 5.32 & 2.32  & 16.7 & 244.51 & -35.04   &  0.06 & 0.52 &   6.98  &  8.85 &  14.15 & 0.80 \\ 
 NGC1904  &   4.352  & -1.57  & -7.86  & -14.12 & 4.00 & 1.72  & 18.8 & 227.23 & -29.35   &  0.16 & 0.80 &   7.78  &  9.10 &  16.23 & 0.90 \\
 NGC2808  &   4.568  & -1.15  & -9.36  & -14.05 & 4.61 & 1.77  & 11.0 & 282.19 & -11.25   &  0.26 & 0.76 &   8.28  &  9.11 &  15.17 & 0.77 \\
 NGC3201  &   4.079  & -1.58  & -7.49  & -15.21 & 2.69 & 1.30  &  9.0 & 277.23 &   8.64   &  1.43 & 2.68 &   8.81  &  9.23 &  18.77 &  0.77 \\
 NGC4147  &   4.061  & -1.83  & -6.16  & -14.26 & 3.48 & 1.80  & 21.3 & 252.85 &  77.19   &  0.10 & 0.43 &   7.49  &  8.67 &  17.63 & 1.03 \\
 NGC4372  &   4.114  & -2.09  & -7.77  & -16.33 & 2.09 & 1.30  &  7.1 & 300.99 &  -9.88   &  1.75 & 3.90 &   8.90  &  9.59 &  20.51 & 0.98 \\  
 NGC4590  &   4.041  & -2.06  & -7.35  & -15.10 & 2.54 & 1.64  & 10.1 & 299.63 &  36.05   &  0.69 & 1.55 &   8.67  &  9.29 &  18.67 & 0.92 \\
 NGC4833  &   4.301  & -1.80  & -8.01  & -15.31 & 3.06 & 1.25  &  6.9 & 303.61 &  -8.01   &  1.00 & 2.41 &   8.71  &  9.34 &  18.45 & 1.01 \\   
 NGC5024  &   4.079  & -1.99  & -8.77  & -14.89 & 3.04 & 1.78  & 18.8 & 332.96 &  79.76   &  0.36 & 1.11 &   8.79  &  9.69 &  17.39 & 1.02 \\   
 NGC5634  &   4.146  & -1.88  & -7.75  & -14.58 & 3.12 & 1.60  & 21.9 & 342.21 &  49.26   &  0.21 & 0.54 &   8.61  &  9.28 &  17.49 & 0.98 \\  
 NGC5694  &   4.204  & -1.86  & -7.81  & -14.13 & 4.03 & 1.84  & 29.1 & 331.06 &  30.36   &  0.06 & 0.33 &   7.86  &  9.15 &  16.34 & 1.05 \\   
 NGC5824  &   4.380  & -1.85  & -8.84  & -13.89 & 4.66 & 2.45  & 25.8 & 332.55 &  22.07   &  0.05 & 0.36 &   7.88  &  9.33 &  15.08 & 1.02 \\   
 NGC5904  &   4.176  & -1.27  & -8.81  & -14.30 & 3.91 & 1.83  &  6.2 &   3.86 &  46.80   &  0.42 & 2.11 &   8.26  &  9.53 &  16.05 & 0.83 \\   
 NGC5927  &   3.724  & -0.37  & -7.80  & -14.75 & 3.87 & 1.60  &  4.5 & 326.60 &   4.86   &  0.42 & 1.15 &   8.29  &  8.98 &  17.45 & 0.94 \\   
 NGC5946  &   4.279  & -1.38  & -7.60  & -14.80 & 4.50 & 2.50  &  7.4 & 327.58 &   4.19   &  0.08 & 0.69 &   7.06  &  8.95 &  17.42 & 0.90 \\   
 NGC5986  &   4.415  & -1.58  & -8.42  & -14.91 & 3.30 & 1.22  &  4.8 & 337.02 &  13.27   &  0.63 & 1.05 &   8.94  &  9.23 &  17.56 & 0.91 \\   
 NGC6093  &   4.477  & -1.75  & -8.23  & -13.68 & 4.76 & 1.95  &  3.8 & 352.67 &  19.46   &  0.15 & 0.65 &   7.73  &  8.86 &  15.19 & 0.97 \\   
 NGC6139  &   4.146  & -1.68  & -8.36  & -15.26 & 4.66 & 1.80  &  3.6 & 342.37 &   6.94   &  0.14 & 0.82 &   7.56  &  9.04 &  17.30 & --- \\   
 NGC6171  &   3.875  & -1.04  & -7.13  & -15.26 & 3.13 & 1.51  &  3.3 &   3.37 &  23.01   &  0.54 & 2.70 &   8.05  &  9.31 &  18.84 & 0.98 \\   
 NGC6205  &   4.505  & -1.54  & -8.70  & -14.53 & 3.33 & 1.51  &  8.7 &  59.01 &  40.91   &  0.78 & 1.49 &   8.80  &  9.30 &  16.80 & 1.05 \\   
 NGC6218  &   4.217  & -1.48  & -7.32  & -15.07 & 3.23 & 1.39  &  4.5 &  15.72 &  26.31   &  0.72 & 2.16 &   8.10  &  9.02 &  18.17 & 0.94 \\   
 NGC6229  &   4.301  & -1.43  & -8.07  & -14.49 & 3.40 & 1.61  & 30.0 &  73.64 &  40.31   &  0.13 & 0.37 &   8.36  &  9.19 &  16.99 & --- \\  
%\hline
%\hline
%\end{tabular}}
%\end{sideways}
%\end{sidewaystable}
%\end{table} 
%\begin{table}[h]
%\footnotesize
%\begin{sideways}
%\scriptsize{\begin{tabular}{||l|c|c|c|c|c|c|c|c|c|c|c|c|c|c||} 
%\hline
%\hline  
%Id&log(Teff$_{HB}$)& $[Fe/H]$& M$_V$$_{t}$&$\Gamma_{col}$&$\rho_{0}$&c&R$_{GC}$&    l    &    b    & r$_{c}$&r$_{h}$& t$_{rc}$ &t$_{rh}$& mu$_{V}$\\      
%\hline  
 NGC6235  &   4.114  & -1.40  & -6.14  & -15.00 & 3.11 & 1.33  &  2.90 & 358.92 &  13.52   &  0.36 & 0.84 &   8.11  &  8.67 &  18.98 & 0.91 \\  
 NGC6266  &   4.477  & -1.29  & -9.19  & -14.22 & 5.14 & 1.70  &  1.70 & 353.58 &   7.32   &  0.18 & 1.23 &   7.64  &  9.19 &  15.35 & 0.92 \\   
 NGC6273  &   4.568  & -1.68  & -9.08  & -14.83 & 3.96 & 1.53  &  1.60 & 356.87 &   9.38   &  0.43 & 1.25 &   8.50  &  9.34 &  16.82 & 0.96 \\   
 NGC6284  &   4.279  & -1.32  & -7.87  & -14.49 & 4.44 & 2.50  &  6.90 & 358.35 &   9.94   &  0.07 & 0.78 &   7.15  &  9.16 &  16.65 & 0.86 \\ 
 NGC6287  &   4.114  & -2.05  & -7.16  & -15.16 & 3.85 & 1.60  &  1.70 &   0.13 &  11.02   &  0.26 & 0.75 &   7.85  &  8.66 &  18.33 & 1.05  \\  
% NGC6293         &   4.176 & -1.920 & -7.770  & ----- & 5.22 & 2.500 &  1.400  357.620    7.830  &  0.05 & 0.910    6.240 &  8.910   16.180 \\   
 NGC6304  &   3.724  & -0.59  & -7.32  & -14.37 & 4.39 & 1.80  &  2.10 & 355.83 &   5.38   &  0.21 & 1.41 &   7.38  &  8.89 &  17.34 & --- \\   
 NGC6342  &   3.778  & -0.65  & -6.44  & -14.64 & 4.77 & 2.50  &  1.70 &   4.90 &   9.73   &  0.05 & 0.88 &   6.09  &  8.66 &  17.44 & 0.92 \\   
 NGC6356  &   3.756  & -0.50  & -8.52  & -14.78 & 3.76 & 1.54  &  7.60 &   6.72 &  10.22   &  0.23 & 0.74 &   8.33  &  9.26 &  17.09 & 0.97 \\  
 NGC6362  &   3.954  & -0.95  & -7.06  & -15.16 & 2.22 & 1.10  &  5.30 & 325.55 & -17.57   &  1.32 & 2.18 &   9.07  &  9.31 &  19.19 & 0.91 \\   
 NGC6388  &   4.255  & -0.60  & -9.82  & -13.96 & 5.31 & 1.70  &  4.40 & 345.56 &  -6.74   &  0.12 & 0.67 &   7.90  &  9.24 &  14.55 & --- \\   
 NGC6397  &   3.978  & -1.95  & -6.63  & -13.96 & 5.68 & 2.50  &  6.00 & 338.17 & -11.96   &  0.05 & 2.33 &   4.90  &  8.46 &  15.65 & 1.00 \\   
 NGC6441  &   4.230  & -0.53  & -9.47  & -14.26 & 5.23 & 1.85  &  3.50 & 353.53 &  -5.01   &  0.11 & 0.64 &   7.72  &  9.13 &  14.99 & --- \\   
% NGC6453         &   4.301 & -1.530 & -7.050  & ----- & 4.72 & 2.500 &  3.300  355.720   -3.870  &  0.07 & 0.370    6.870 &  8.360   17.350 \\   
% NGC6522         &   4.342 & -1.440 & -7.670  & ----- & 5.31 & 2.500 &  0.600    1.020   -3.930  &  0.05 & 1.040    6.320 &  8.900   16.140 \\   
% NGC6540         &   4.041 & -1.200 & -5.380  & ----- & 5.92 & 2.500 &  4.400    3.290   -3.310  &  0.03 & 0.240    5.010 &  7.080   16.400 \\   
 NGC6544  &   4.176  & -1.56  & -6.56  & -14.81 & 5.75 & 1.63  &  5.40 &   5.84 &  -2.20   &  0.05 & 1.77 &   5.05  &  8.35 &  17.13 & 0.84 \\   
 NGC6569  &   3.954  & -0.86  & -7.68  & -15.25 & 2.92 & 1.20  &  7.00 & 342.14 & -16.41   &  0.59 & 0.80 &   9.01  &  9.09 &  17.79 & --- \\   
 NGC6624  &   3.771  & -0.44  & -7.50  & -13.74 & 5.25 & 2.50  &  1.20 &   2.79 &  -7.91   &  0.06 & 0.82 &   6.62  &  8.74 &  15.42 & 0.88 \\   
 NGC6637  &   3.748  & -0.70  & -7.52  & -14.27 & 3.81 & 1.39  &  1.60 &   1.72 & -10.27   &  0.34 & 0.83 &   8.15  &  8.79 &  16.83 & 0.89 \\   
 NGC6638  &   4.097  & -0.99  & -6.83  & -14.42 & 4.05 & 1.40  &  1.60 &   7.90 &  -7.15   &  0.26 & 0.66 &   7.93  &  8.51 &  17.27 & 0.87 \\   
 NGC6642  &   4.061  & -1.35  & -6.57  & -14.16 & 4.72 & 1.99  &  1.60 &   9.81 &  -6.44   &  0.10 & 0.73 &   6.94  &  8.49 &  16.68 & --- \\   
 NGC6652  &   4.000  & -0.96  & -6.57  & -13.92 & 4.54 & 1.80  &  2.40 &   1.53 & -11.38   &  0.07 & 0.65 &   6.66  &  8.55 &  16.31 & 0.89 \\   
 NGC6681  &   4.301  & -1.51  & -7.11  & -13.73 & 5.41 & 2.50  &  2.10 &   2.85 & -12.51   &  0.03 & 0.93 &   5.62  &  8.83 &  15.28 & 0.93 \\   
% NGC6712         &   4.021 & -1.010 & -7.500  & ----- & 3.14 & 0.900 &  3.500   25.350   -4.320  &  0.94 & 1.370    8.860 &  8.980   18.650 \\   
 NGC6717  &   4.114  & -1.29  & -5.67  & -13.76 & 4.65 & 2.07  &  2.30 &  12.88 & -10.90   &  0.94 & 1.37 &   6.61  &  8.26 &  16.48 & 0.92 \\   
 NGC6723  &   4.130  & -1.12  & -7.86  & -14.82 & 2.81 & 1.05  &  2.60 &   0.07 & -17.30   &  0.94 & 1.61 &   8.99  &  9.30 &  17.92 & 0.96 \\   
% NGC6760         &   3.740 & -0.520 & -7.860  & ----- & 3.84 & 1.590 &  4.800   36.110   -3.920  &  0.33 & 2.180    7.940 &  9.390   18.790 \\   
 NGC6838  &   3.763  & -0.73  & -5.56  & -14.92 & 3.05 & 1.15  &  6.70 &  56.74 &  -4.56   &  0.63 & 1.65 &   7.64  &  8.41 &  19.22 & 0.91 \\  
 NGC6864  &   4.176  & -1.16  & -8.35  & -13.98 & 4.51 & 1.88  & 12.80 &  20.30 & -25.75   &  0.10 & 0.47 &   7.85  &  9.08 &  15.55 & 0.85 \\   
 NGC6934  &   4.130  & -1.54  & -7.65  & -14.45 & 3.37 & 1.53  & 14.30 &  52.10 & -18.89   &  0.25 & 0.60 &   8.43  &  9.07 &  17.26 & 0.85 \\   
 NGC6981  &   4.000  & -1.40  & -7.04  & -15.07 & 2.35 & 1.23  & 12.90 &  35.16 & -32.68   &  0.54 & 0.88 &   8.93  &  9.20 &  18.90 & 0.83 \\   
 NGC7078  &   4.477  & -2.26  & -9.17  & -13.65 & 5.38 & 2.50  & 10.40 &  65.01 & -27.31   &  0.07 & 1.06 &   7.02  &  9.35 &  14.21 & 0.94 \\   
 NGC7089  &   4.477  & -1.62  & -9.02  & -14.22 & 3.90 & 1.80  & 10.40 &  53.38 & -35.78   &  0.34 & 0.93 &   8.54  &  9.32 &  15.92 & 0.94 \\   
 NGC7099  &   4.079  & -2.12  & -7.43  & -13.72 & 5.04 & 2.50  &  7.10 &  27.18 & -46.83   &  0.06 & 1.15 &   6.38  &  8.95 &  15.28 & 1.08 \\  
\hline
\hline
\end{tabular}
\centering
\caption[Funciona]{Col. 1: cluster  identification; Col.  2: logarithm of HB highest effective
temperature; Col.  3: metallicity; Col. 4: total cluster luminosity in $V$ or absolute
visual  magnitude;  Col.  5: collisional parameter; Col.  6: logarithm    of
central  luminosity   density; Col. 7  central concentration;  Col.  8, 9
and 10: distance from Galactic  center,  Galactic
longitude and latitude (degrees); Col.  11 and 12: core radius and
half-light radius in  arcmin; Col. 13 and 14: logarithm of
core relaxation  time,   and  logarithm of   relaxation   time at  the
half-light radius, in  log$_{10}$ (years); Col. 15:
central surface brightness in V magnitudes per square arcsecond; Col. 16: Cluster relative age.}}
%\end{sideways}
\end{table*}

\subsection{Other parameters}
\label{param}

In order to disentangle the dependence of the HB morphology on as many
cluster  parameters as  possible, we  have  considered  the  15
quantities listed in Table 1 : maximum effective temperature along
the HB (log(Teff$_{HB}$)); cluster metallicity ($[Fe/H]$);
total  luminosity (M$_V$);  collisional   parameter ($\Gamma_{col}$);  
logarithm   of   central luminosity  density in   Solar
luminosities per cubic parsec ($\rho_{0}$);   central 
concentration (c$=$log(r$_t$/r$_c$));  distance  from Galactic 
center in kpc assuming R$_{0}$=8.0  kpc (R$_{GC}$);  Galactic   longitude (l) and   latitude (b);  core  radius
(r$_{c}$); half-light radius (r$_{h}$); the logarithm  of    core relaxation 
time (t$_{rc}$); the   logarithm  of relaxation time    at    half-light radius 
(t$_{rh}$); the    central  surface brightness ($\mu_{V}$) and the age in a
relative scale (Age$_{rel}$).

The  collisional parameter, listed  in
column 5, is defined as the probability of collisions, per unit time, 
for one star in the cluster, and it
was derivated via the formula (King, 2002):

\begin{center}

$\Gamma_{col} = (\log [5 \cdot 10^{-15} \sqrt{\sigma^3 \cdot r_c}])/N_{star}$

\end{center}

where r$_c$ is the core radius in units of parsecs and $\sigma$  is  
the  central   surface brightness  in  units   of L$_\odot$/pc$^2$:
\begin{center}
  $\sigma = 10^{[-0.4 \cdot (\mu_V - 26.41)]}$
\end{center}

The total number of stars in the cluster, N$_{star}$  has been
estimated by using the integrated visual absolute magnitude of the cluster,
assuming $M/L=2$ and a typical mass for the colliding stars of 0.4 $m_{\sun}$.

Column 16 gives the cluster relative ages from a subsample of 47
clusters in  common with De Angeli et al.\ (2005). They
used the so-called vertical method to estimate ages for a sub-sample of 41
cluster from the snapshot database and 30 clusters from the ground based
database presented in Rosenberg et al. (2000a, 2000b). Sixteen clusters
were in common among the two database and were used to assess the
consistency of the two catalogs.
Our analysis will include 3 clusters from the ground based catalog and 39
clusters from the snapshot one. Five more snapshot clusters had age
estimates that did not match the accuracy of the other determinations and
for this reason were not included in the final version of De Angeli et al.\ (2005),
although their ages had been determined homogeneously with respect to the
rest of the catalog. Nevertheless, we decided to include them in our
analysis given their statistical value.

The other quantities in Table 1 (see table's caption) are taken from
Harris (1996, in its revised version of 2003).

\section{Monovariate correlations}

We  first explore the   simple,  pairwise correlations between  the HB
extension  and a number of  selected globular  cluster properties, via
the  Pearson coefficient,  r.  This coefficient gives  the ratio  between the observed covariance
and the  maximum  possible positive covariance  for the  two evaluated
quantities, $x$ and $y$  .  Therefore, the value  of
r goes from perfect negative  correlation (r$=-1$) to perfect
positive  correlation (r$=+1$).   The  midpoint of   this range,  r$=0.0$,
corresponds to a complete absence of correlation. 

The  coefficient  of  determination,   r$^2$,  represents the
correlation strength. The value   of r$^2 \cdot 100$ is   the
percentage  of   variability    in $y$    associated  with
variability in $x$. 

Table 2 lists  the  Pearson    linear   regression  correlation
coefficients, r, for the 14 quantities explored. Columns in
Table 2 list the same quantities as in Table 1. In addition,
the same correlation matrix has been computed for intermediate
metallicity clusters only (-1.8 $<$ [Fe/H] $<$ -1.3, Table 3). This allows to analyse the
impact of the various correlations on a metallicity regime
with a high sensitivity to any variation in the basic stellar parameters.

%\begin{sidewaystable}[h]
\begin{table*}
\centering
\scriptsize{\begin{tabular}{lcccccccccccccc} 
\hline
\hline  
       &log(Teff$_{HB}$)& $[Fe/H]$& M$_V$ & $\Gamma_{col}$&$\rho_{0}$&c&R$_{GC}$& {\it l}   &{\it b}& r$_{c}$&r$_{h}$& t$_{rc}$ &t$_{rh}$& $\mu_{V}$\\   
\hline
log(Teff$_{HB}$)& 1.00 &-0.54 &-0.48 & 0.14 & 0.14 & 0.09 & 0.22 & 0.22 & 0.09 &-0.07 &-0.19 & 0.11 & 0.28 &-0.31\\
$[Fe/H]$        &-0.54 & 1.00 & 0.00 & 0.10 & 0.16 &-0.10 &-0.40 &-0.16 &-0.21 &-0.10 &-0.08 &-0.03 &-0.19 &-0.08\\
M$_V$           &-0.48 & 0.00 & 1.00 &-0.22 &-0.18 &-0.08 &-0.17 &-0.33 & 0.06 & 0.11 & 0.08 &-0.37 &-0.73 & 0.53\\
$\Gamma_{col}$  & 0.14 & 0.10 &-0.22 & 1.00 & 0.76 & 0.66 & 0.13 &-0.03 &-0.28 &-0.66 &-0.51 &-0.49 & 0.28 &-0.91\\
$\rho_{0}$      & 0.14 & 0.16 &-0.18 & 0.76 & 1.00 & 0.77 &-0.19 &-0.07 &-0.25 &-0.72 &-0.32 &-0.80 &-0.41 &-0.83\\
c               & 0.09 &-0.10 & 0.08 & 0.66 & 0.77 & 1.00 & 0.06 & 0.00 &-0.07 &-0.62 &-0.27 &-0.72 &-0.22 &-0.68\\
R$_{GC}$        & 0.22 &-0.41 &-0.18 & 0.13 &-0.19 & 0.06 & 1.00 & 0.24 & 0.25 &-0.17 &-0.32 & 0.22 & 0.32 &-0.09\\
{\it l}         & 0.22 &-0.16 &-0.33 &-0.03 &-0.07 & 0.00 & 0.24 & 1.00 & 0.14 & 0.08 &-0.03 & 0.24 & 0.32 &-0.05\\
{\it b}         & 0.09 &-0.21 & 0.06 &-0.28 &-0.25 &-0.07 & 0.25 & 0.14 & 1.00 & 0.00 &-0.04 & 0.16 & 0.14 & 0.27\\
r$_{c}$         &-0.07 &-0.10 & 0.11 &-0.66 &-0.72 &-0.62 &-0.17 & 0.08 & 0.00 & 1.00 & 0.73 & 0.56 & 0.33 & 0.65\\
r$_{h}$         &-0.19 &-0.08 & 0.08 &-0.51 &-0.32 &-0.27 &-0.32 & 0.03 &-0.04 & 0.73 & 1.00 & 0.12 & 0.26 & 0.42\\
t$_{rc}$        & 0.11 & 0.02 &-0.37 &-0.49 &-0.80 &-0.72 & 0.22 & 0.24 & 0.16 & 0.56 & 0.12 & 1.00 & 0.68 & 0.43\\
t$_{rh}$        & 0.28 &-0.19 &-0.73 & 0.28 &-0.42 &-0.22 & 0.32 & 0.32 & 0.15 & 0.33 & 0.26 & 0.68 & 1.00 & 0.05\\
$\mu_{V}$        &-0.31 &-0.08 & 0.53 &-0.91 &-0.84 &-0.68 &-0.09 &-0.05 & 0.27 & 0.65 & 0.42 & 0.43 & 0.05 & 1.00\\
\hline
\hline
\end{tabular}}
\caption{Pearson  linear    regression  correlation
coefficients, $\textbf{r}$, for the 14 quantities in Table 1.}
%\end{sidewaystable}
%\end{sideways}
\end{table*}

\begin{table*}
\centering
\scriptsize{\begin{tabular}{lcccccccccccccc} 
\hline
\hline  
       &log(Teff$_{HB}$)& $[Fe/H]$& M$_V$ & $\Gamma_{col}$&$\rho_{0}$&c&R$_{GC}$& {\it l}   &{\it b}& r$_{c}$&r$_{h}$& t$_{rc}$ &t$_{rh}$& $\mu_{V}$\\   
\hline
log(Teff$_{HB}$)&  1.00 & -0.57 & -0.77 &  0.29 &  0.30 &  0.26 & -0.11 &  0.16 &  0.38 & -0.11 & -0.02 &  0.02 &  0.36 & -0.60 \\
$[Fe/H]$        & -0.57 &  1.00 &  0.50 & -0.03 & -0.05 &  0.24 &  0.20 & -0.16 & -0.10 & -0.35 & -0.43 & -0.16 & -0.27 &  0.22 \\
M$_V$           & -0.77 &  0.50 &  1.00 & -0.13 &  0.15 &  0.04 & -0.19 & -0.24 & -0.10 & -0.18 &  0.03 & -0.49 & -0.78 &  0.37 \\
$\Gamma_{col}$  &  0.29 & -0.03 & -0.13 &  1.00 &  0.58 &  0.58 &  0.02 & -0.19 & -0.12 & -0.64 & -0.61 & -0.37 & -0.25 & -0.90 \\
$\rho_{0}$      &  0.30 & -0.05 &  0.15 &  0.58 &  1.00 &  0.76 & -0.35 & -0.20 &  0.07 & -0.70 & -0.20 & -0.91 & -0.64 & -0.73 \\
c               &  0.26 &  0.24 &  0.04 &  0.58 &  0.76 &  1.00 & -0.19 & -0.07 &  0.08 & -0.66 & -0.37 & -0.68 & -0.31 & -0.66 \\
R$_{GC}$        & -0.11 &  0.20 & -0.19 &  0.02 & -0.35 & -0.19 &  1.00 & -0.18 & -0.14 & -0.08 & -0.31 &  0.31 &  0.36 &  0.04 \\
{\it l}         &  0.16 & -0.16 & -0.24 & -0.19 & -0.20 & -0.07 & -0.18 &  1.00 &  0.07 &  0.18 & -0.03 &  0.32 &  0.27 &  0.17 \\
{\it b}         &  0.38 & -0.10 & -0.10 & -0.12 &  0.07 &  0.08 & -0.14 &  0.07 &  1.00 &  0.13 &  0.20 & -0.02 & -0.04 & -0.00 \\
r$_{c}$         & -0.11 & -0.35 & -0.18 & -0.64 & -0.70 & -0.66 & -0.08 &  0.18 &  0.13 &  1.00 &  0.77 &  0.63 &  0.52 &  0.62 \\
r$_{h}$         & -0.02 & -0.43 &  0.03 & -0.61 & -0.15 & -0.37 & -0.31 & -0.03 &  0.20 &  0.77 &  1.00 &  0.07 &  0.14 &  0.43 \\
t$_{rc}$        &  0.02 & -0.16 & -0.49 & -0.37 & -0.91 & -0.68 &  0.31 &  0.32 & -0.02 &  0.63 &  0.07 &  1.00 &  0.77 &  0.45 \\
t$_{rh}$        &  0.36 & -0.27 & -0.78 & -0.25 & -0.64 & -0.31 &  0.36 &  0.27 & -0.04 &  0.52 &  0.14 &  0.77 &  1.00 &  0.13 \\
$\mu_{V}$        & -0.60 &  0.22 &  0.37 & -0.90 & -0.73 & -0.66 &  0.04 &  0.17 & -0.00 &  0.62 &  0.43 &  0.45 &  0.13 &  1.00 \\
\hline
\hline
\end{tabular}}
\caption{Same as Table 2, but for intermediate metallicity clusters only.}
%\end{sidewaystable}
%\end{sideways}
\end{table*}

We note that among  the parameters studied here,  only 8 are  measured
independently.    $\rho_{0}$  is  derived  from    $\mu_{V}$(0), c and
r$_{c}$; t$_{rh}$ is  derived from the M$_{V}$ and  r$_{h}$,  etc.  In
general, correlations  of any  of the derived  quantities  with any of
their constituent quantities   or combinations do not provide new information.
On the other   hand,   Tables 2 and 3 immediately suggest  some
interesting correlations that we will try to analyse next.

\begin{itemize}

\item \textbf{Metallicity: the first parameter}

The first correlation to be explored  is the HB morphology-metallicity
dependence.  As  pointed out in the Introduction,  metallicity  is the so called
first  parameter    regulating   the  extension   of   the  horizontal
branch, and its influence can be  naturally derived   from canonical
stellar   evolution   models.  Figure    3   shows  the   trend   of
log(Teff$_{HB}$) with  $[Fe/H]$. Clearly, there   is a correlation between
both quantities in  the sense that  the less metallic the  cluster is,
the more  extended  its HB tends to  be.    However,  the  data  indicate, 
as   we  already  knew, that the variation of   log(Teff$_{HB}$) from    
cluster  to cluster    is  not
completely explained   by  the  $[Fe/H]$  parameter. This  observational
evidence  is  the   core of  the    second parameter problem, mentioned in the
Introduction.   
The  value of  the  Pearson  correlation
coefficient  for these two quantities  is r $\simeq$ -0.54 (and r $\simeq$ -0.57
in the intermediate metallicity regime), indicating
that metallicity   explains  the $\simeq$ 30\%   of  the total   variation   of
log(Teff$_{HB}$). A simple least-square fit is  also plotted in Figure
3, giving a rms value of 0.19.

The influence of the selected metallicity scale on the result has been analysed
by repeating the calculations in the Carretta \& Gratton (1997) metallicity scale.
The value of the derived correlation coefficient between $[Fe/H]$  and log(Teff$_{HB}$)
slightly diminishes (r $\simeq$ -0.47).

\begin{figure*}
\centering{\includegraphics[width= 10 cm]{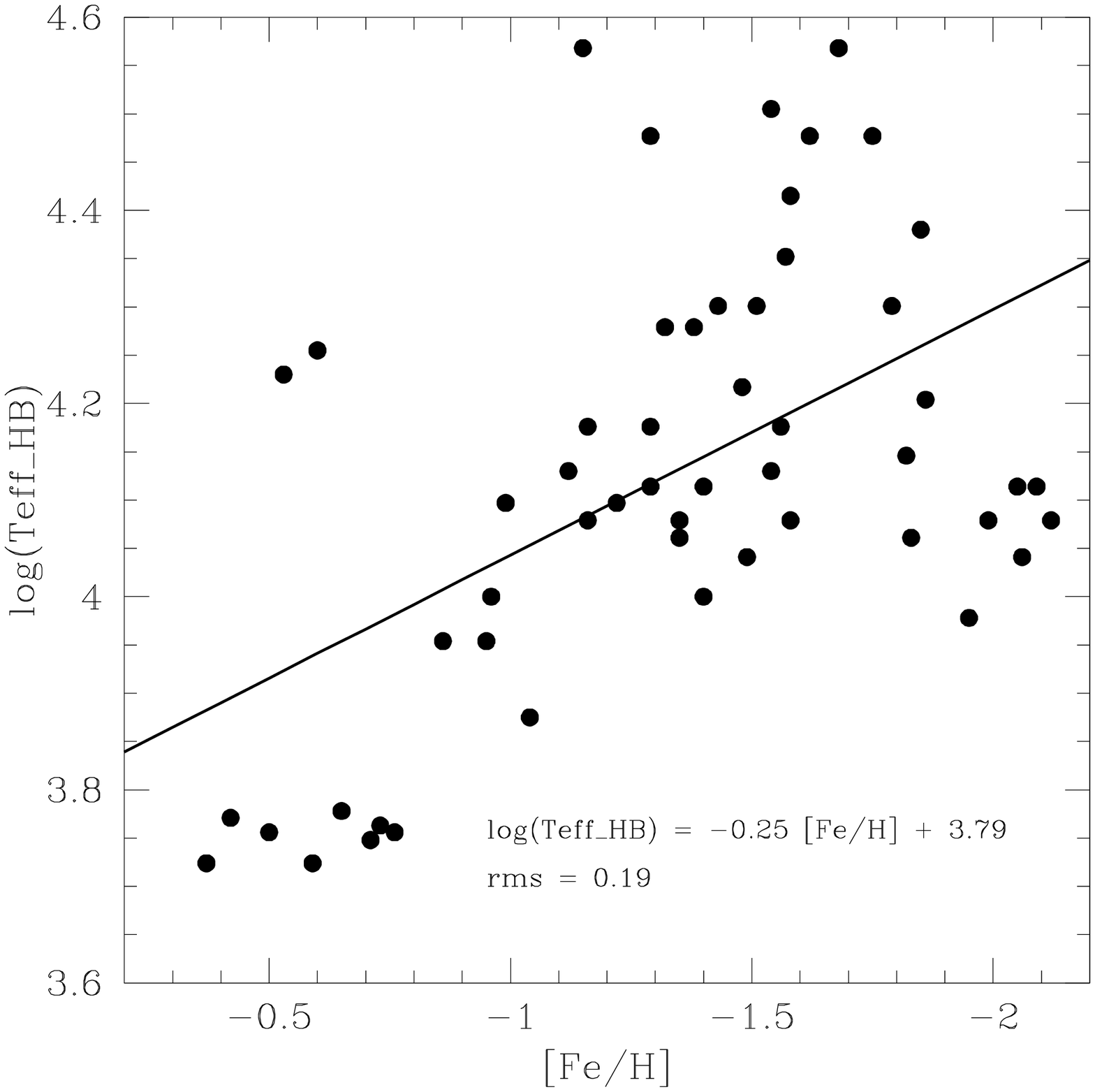}} 
\protect\caption[]
{Correlation of HB morphology with metallicity (the first parameter).
HB morphology is parameterized via the highest effective
temperature reached in the HB, log(Teff$_{HB}$).}
\end{figure*}

\begin{figure*}
\begin{center}
\begin{tabular}{c}
\includegraphics[width= 10 cm]{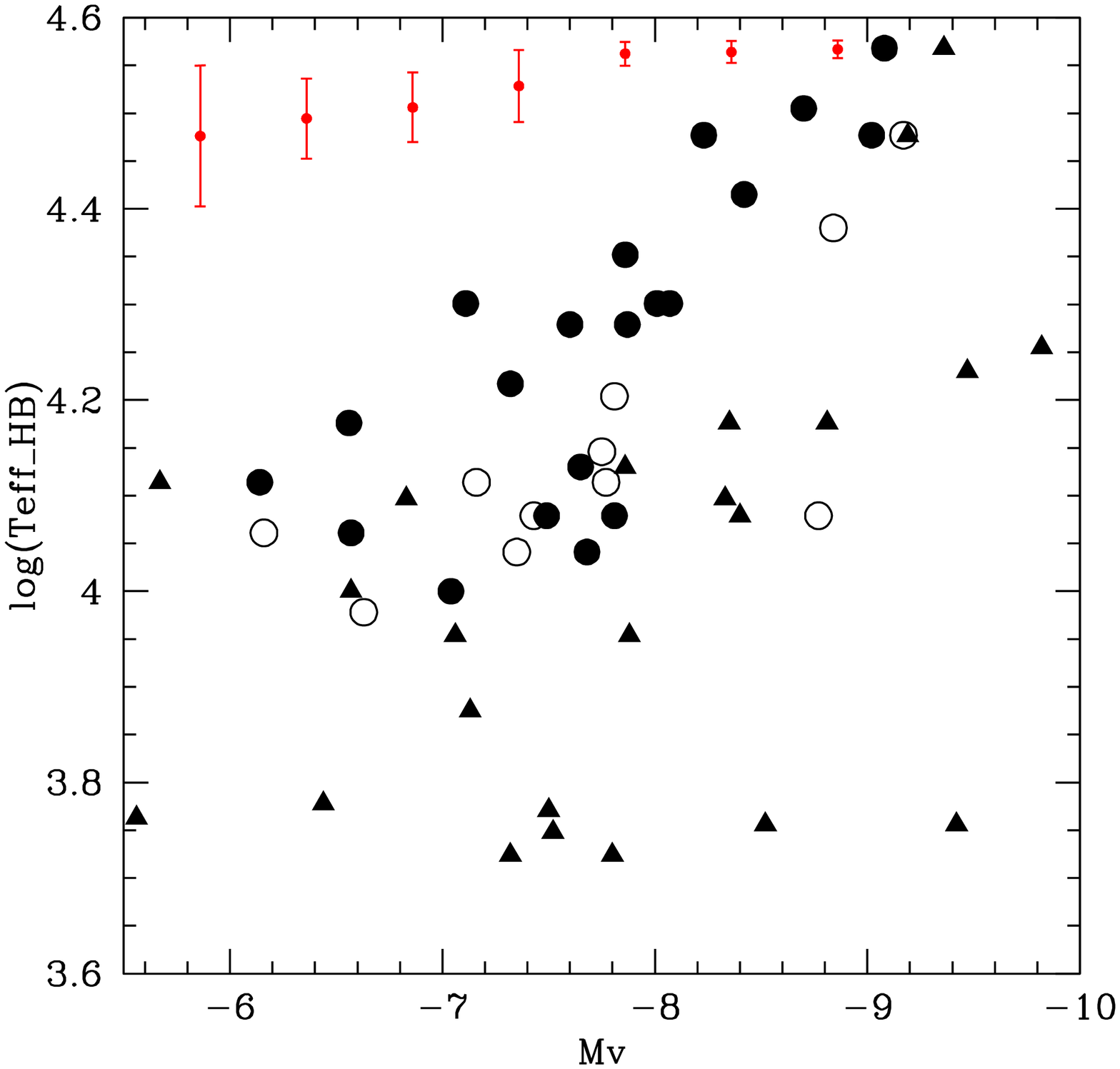} \\
\includegraphics[width= 10 cm]{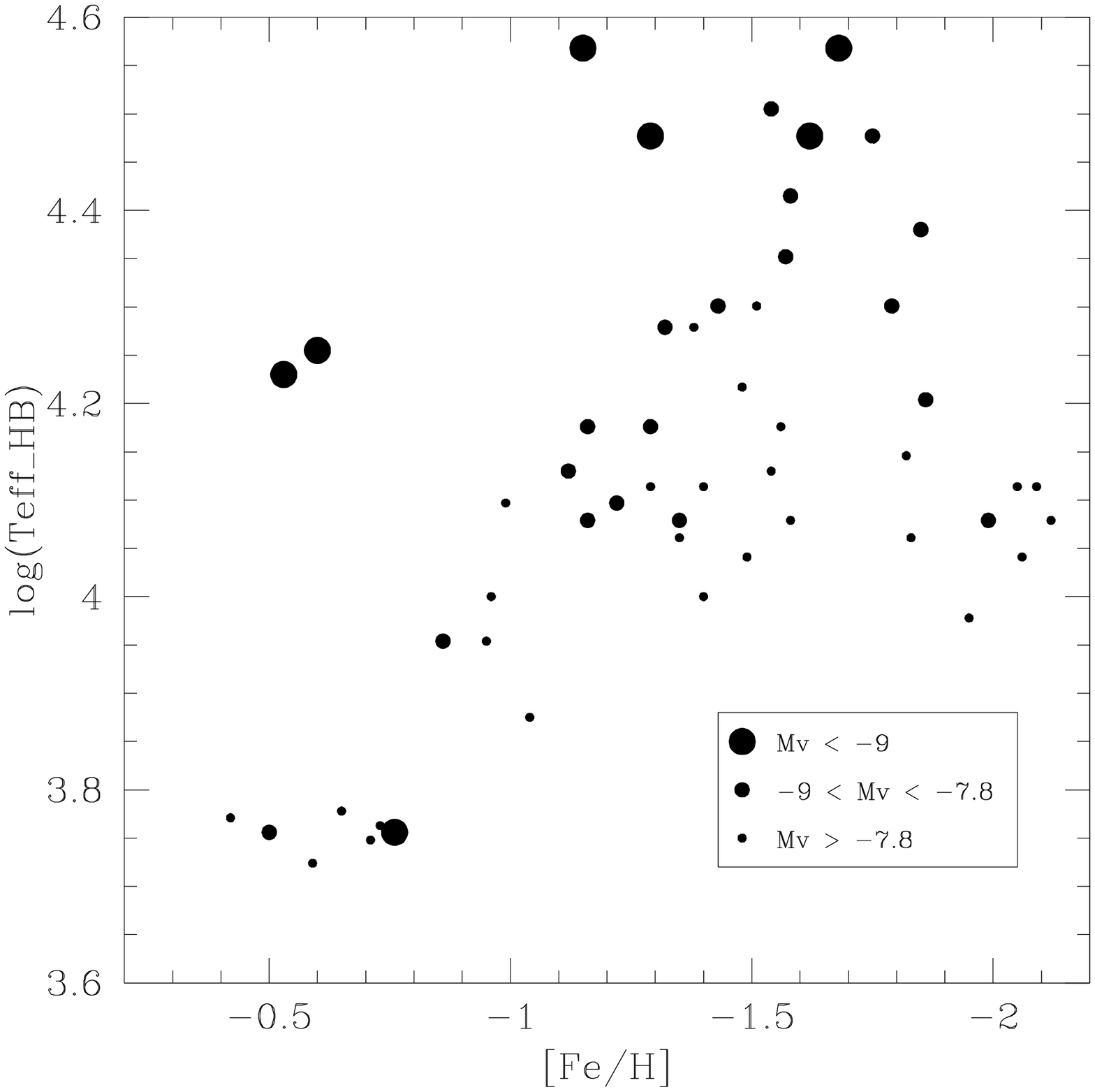} \\
\end {tabular}
\end{center}
\protect\caption[]
{Upper panel:  correlation of log(Teff$_{HB}$)  with total luminosity,
for 3  different metallicity intervals: $[Fe/H] <-1.8$ (open circles),
$-1.8 < [Fe/H] < -1.3$ (filled circles) and $[Fe/H] > -1.3$ (filled triangles).   
Points with error bars are measurements from simulated clusters (e.g. the text).  
Bottom   panel: HB  morphology-metallicity correlation  with
different symbol sizes depending on total  luminosity, M$_V$. Note the
better  correlation    of clusters of  intermediate    luminosity. The
smallest clusters  (  M$_V  >$ -7.8  )  do not  reach the   bluest  HB
morphologies, and they all have log(Teff$_{HB}$) $<$ 4.35. }
%\label{firstp}
\end{figure*}

%\begin{figure}
%\begin{center}
%\begin{tabular}{c}
%\includegraphics[width= 6.7 cm]{Nhb_mv.ps}
%\includegraphics[width= 6.7 cm]{Nbhb_teff.ps}
%\end {tabular}
%\end{center}
%\protect\caption[]
%{The number of HB stars hotter than the HB turning down (in the F439W-F555W,F555W 
%plane) relative to the total flux in the same region, 
%gainst M$_v$ (left panel) and the maximum temperature extension of the HB (right
%panel). The vertical line marks the
%temperature limit between blue HB stars and extreme HB stars. Starred points are 
%EHB clusters in both panels.}
%\label{Nhb}
%\end{figure}

\item \textbf{Total luminosity}

Probably   one  of  the  most   interesting   results  of  this simple
correlation approach is the finding  of a clear correlation between HB
morphology  and total  luminosity  of the  cluster. If no selection
in metallicity is performed, this correlation is
apparently slightly lower than   the  one  observed with   the    first parameter.  The  Pearson
coefficient   relating  these two   variables  is   r  $\simeq$ -0.48
and therefore, the variation of total  luminosity would be responsible for  the 23\%
of  the HB morphology variation. However, if only intermediate metallicity
clusters are considered, the correlation between   log(Teff$_{HB}$) and  M$_V$
is as high as r  $\simeq$ -0.77 (60 \% of the total variation). Therefore, the
influence of cluster total luminosity on the HB temperature extension seems more
important than that of [Fe/H] at this metallicity regime! 

Graphically,  this effect is shown in
Figure 4, upper panel,  where the correlation of log(Teff$_{HB}$)
with  total luminosity  for  3   different  metallicity intervals   is
presented.   More luminous clusters,   tend  to have hotter  (bluer)
horizontal  branches. The rms  of a linear regression between both quantities is 0.20. 
 %Interestingly,  the   highest  correlation  is
%observed for intermediate metallicity clusters  (full circles) where the
%first parameter  effect is less important.   
On the other hand,  if we
calculate  the  Pearson  coefficient  between   total  luminosity  and
log(Teff$_{HB}$)  only for the  subsample  of  clusters with the  more
extended  HBs, log(Teff$_{HB}$)  $>$ 4.3,  we  find a much clearer
correlation, reaching a   value of r$=$ -0.81. This means that two thirds
of the variation in the HB temperature extension can be explained by the
variation of M$_V$ for clusters with extreme blue horizontal branches.   
Bottom  panel of
Figure  4  shows    again  the  trend  of    log(Teff$_{HB}$)  with
metallicity, but using different point sizes depending on the value of the
total   luminosity.   Clearly, some  of  the  dispersion  in the first
parameter correlation can be explained by M$_V$.  That  is the case of
the   metal-rich blue tail  clusters  NGC6388  and NGC6441, which have
M$_V$ $<$ -9, and are among the most luminous clusters in the sample.

Luminosity   is   perhaps the    most   fundamental  observed quantity
characterizing a stellar system, and for  a set of old stellar systems
it is a  good relative measure   of its baryonic mass.  Therefore, the
observed trend  seems to suggest  that, for  some reason, more massive
clusters tend to have bluest horizontal branches. We will come back to
this   point in  Section 7,  where   we  will  discuss its  possible
theoretical implications. 

On the other hand, correlation between HB morphology and total luminosity was previously
noted by Fusi Pecci et al. (1993). However, they interpreted this result
as a consequence of the high HB morphology-cluster density correlation
derived from their analysis. In this paper, we have chosen a different
characterization of cluster HB, the maximum temperature extension, which shows
a weak correlation with cluster density or even stellar collisions,
as we will see later in this Section. 

Finally, we have explored the possibility that the Teff$_{HB}$-M$_V$
dependence could be a statistical effect, that is, the higher 
the number of stars in the cluster (and thus the more massive the cluster is)
the higher the probability of finding hot HB stars. 
The so called ``second parameter'' could also  be a mechanism with a low percentage
of incidence, that would be only detected with high enough statistics.
In principle, if this were true, we would expect the hot HB stars to be always a
small percentage of the total number of HB stars, which is not always the case
(see for example the cluster NGC6205).  Nevertheless, in order to check the
influence of statistics on the M$_V$ parameter dependence, we have performed
the following test: we have taken the photometry of one of the most massive 
clusters in the sample, NGC~2808 (M$_V= -9.36$), whose extremely extended HB reaches a
temperature of log(Teff$_{HB}$)$=$4.568.  The absolute total magnitude of
the cluster was then reduced artificially by subtracting the corresponding
percentage of stars from the photometry file, using a random selection procedure.
Between the 34\% and the 96\% of the stars were removed for magnitude reductions between
0.5 mag and 3.5 mag.
From the resulting simulated CMDs, the highest temperature of the HB was measured following
the same technique applied to the real clusters (see Section 3.1). This
procedure was repeated 20 times for each simulated cluster magnitude.
The results obtained are plotted in Figure 4, upper panel, where the points with error
bars correspond to the mean log(Teff$_{HB}$) value obtained for each cluster magnitude
and its scatter. From our simulations, the HB temperature extension seems to
decrease very little as cluster magnitude decreases (less than a 4\% in all the
magnitude range). On the contrary, the tendency for real clusters (points without
error bars in Fig. 4), seems to indicate a steeper decrease of log(Teff$_{HB}$)
with M$_V$. Therefore, the performed test proves that the role of statistics in
the dependence of HB temperature extension on cluster magnitude is very small.
The origin of the log(Teff$_{HB}$)-M$_V$ correlation could be a different physical
cause, whose possible interpretation will be discussed later (cf. Section 7).

\item \textbf{Collisional parameter}

\begin{figure*}
\begin{center}
\begin{tabular}{cc}
\includegraphics[width= 6.5 cm]{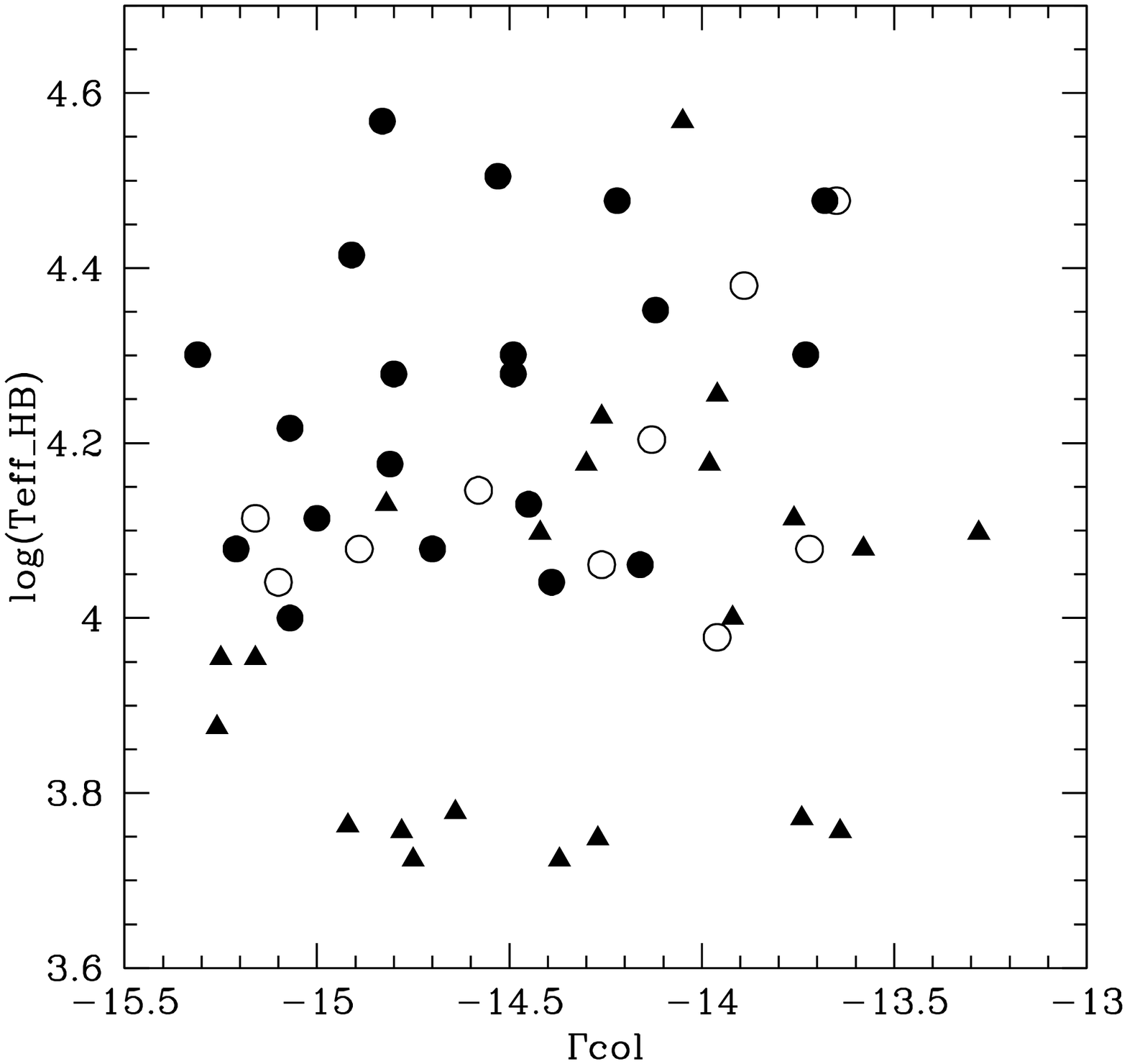}&\includegraphics[width= 6.5 cm]{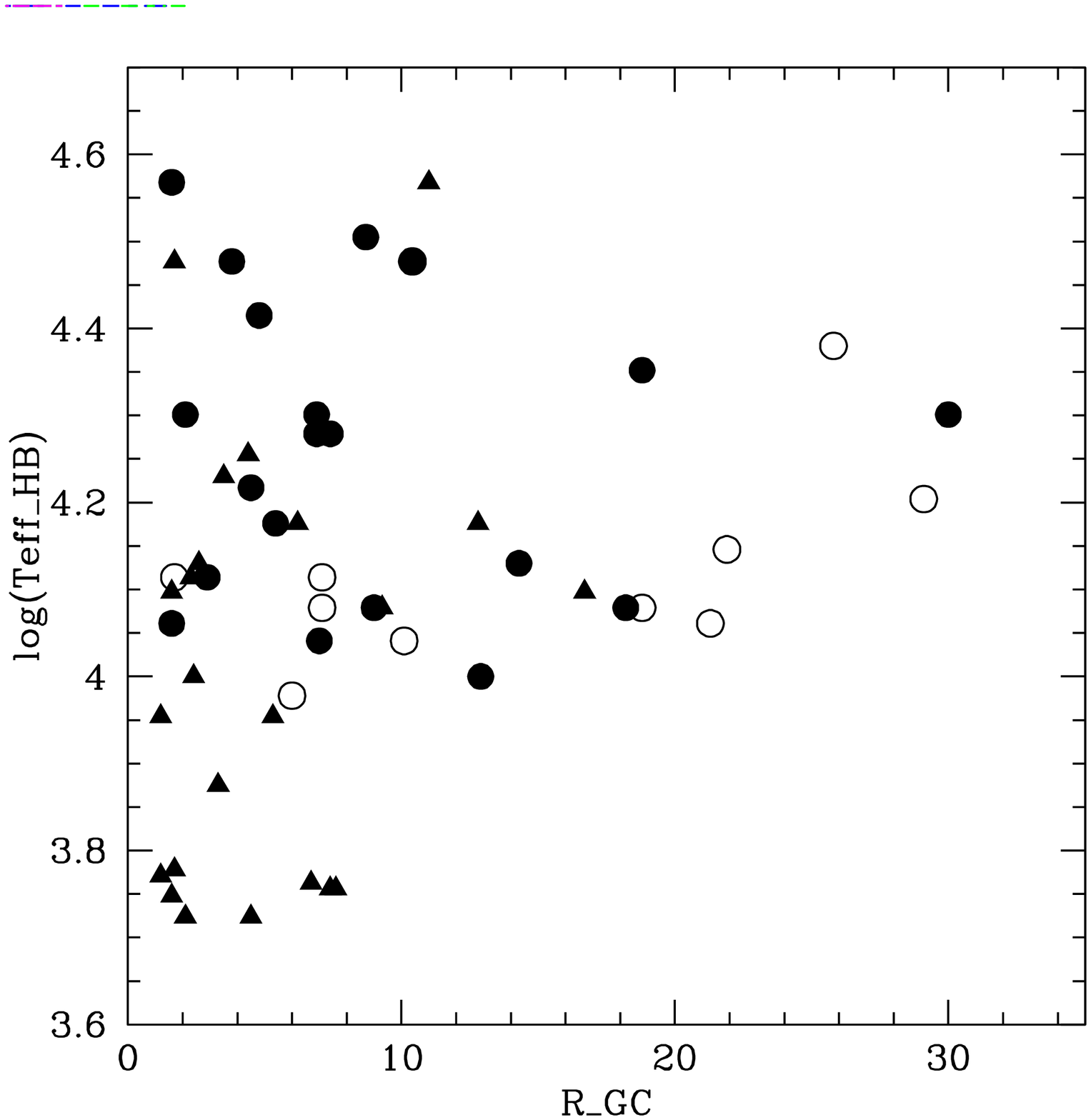}\\
\includegraphics[width= 6.5 cm]{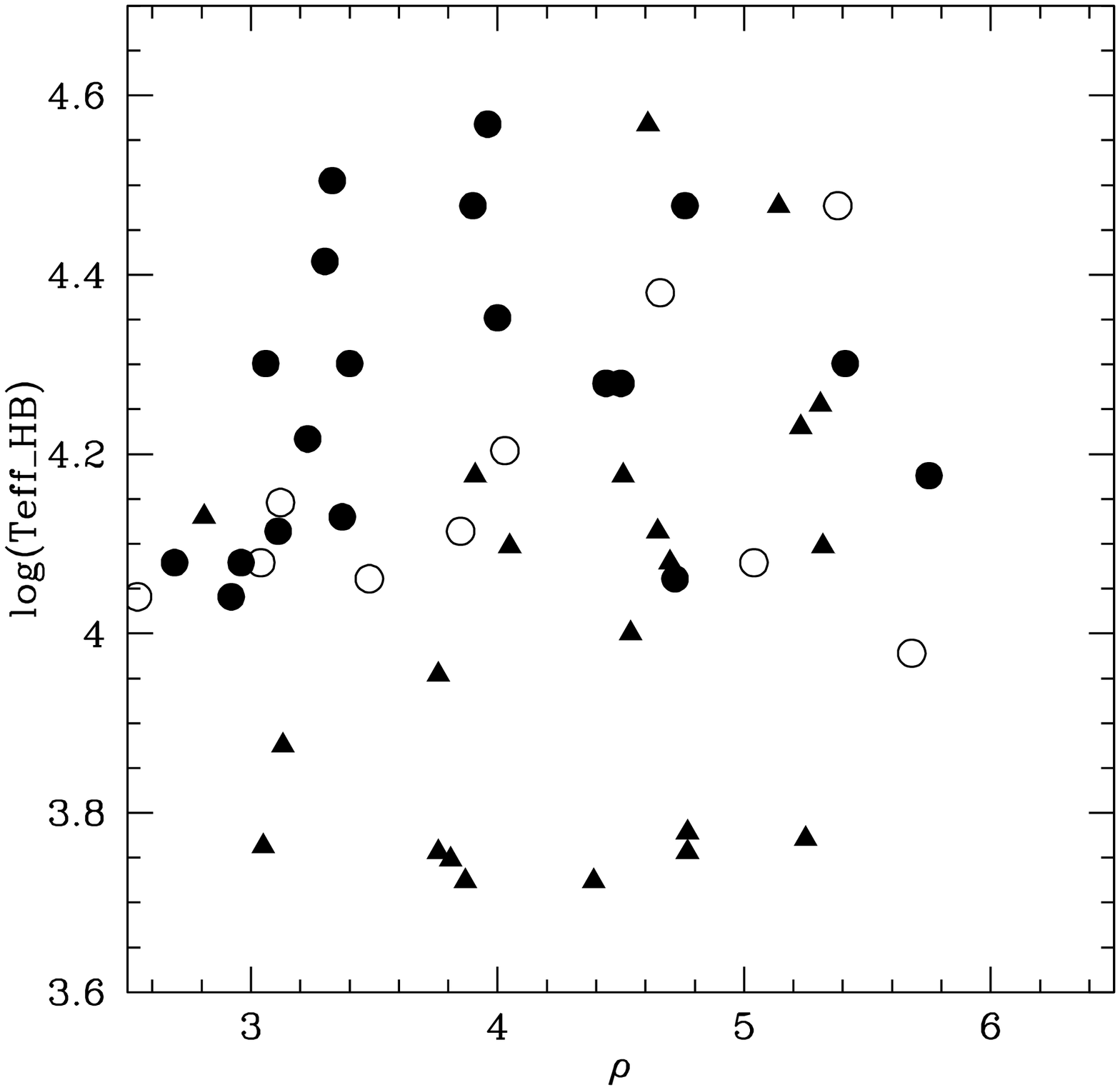}& \\
\end{tabular}
\end{center}
\protect\caption[]
{Correlation  of log(Teff$_{HB}$) with    (from upper left to   bottom
right) collisional parameter, central density and Galactocentric distance. 
Symbols indicate different $[Fe/H]$ ranges, as in Figure 4.}
%\label{firstp}
\end{figure*}

\begin{figure*}
\centerline{\includegraphics[width= 10 cm]{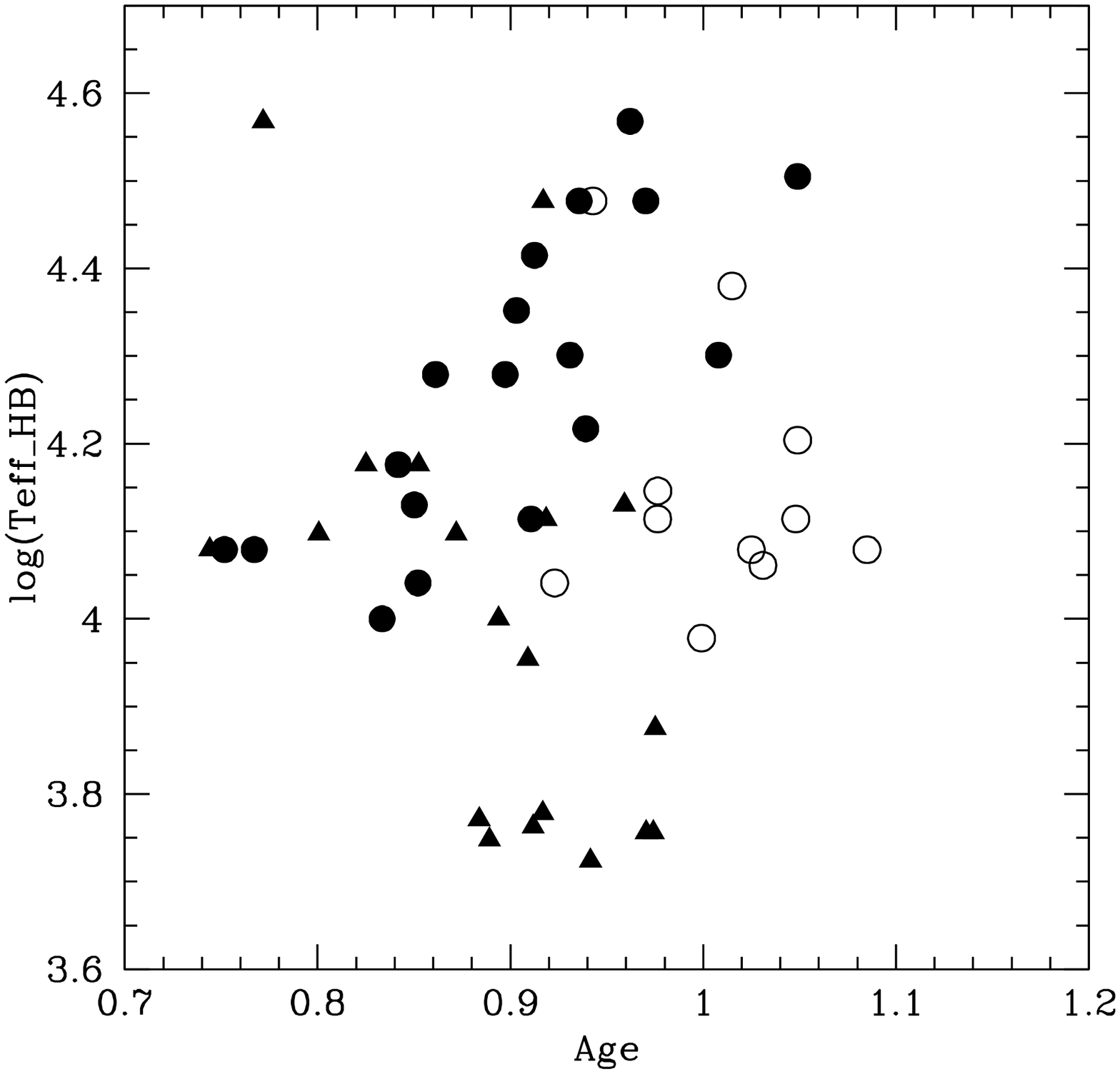}}
\protect\caption[]
{Correlation of log(Teff$_{HB}$) with globular cluster relative age. 
Symbols are the same as in Fig. 4
The derived  Pearson  correlation  coefficient
is  r$=$ 0.04 when all the clusters are considered, but it increases
up to r$=$0.76 if only intermediate clusters (filled circles) are taken into account. }
\label{firstp}
\end{figure*}

Quite interesting is  also the result for the collisional parameter.
The correlation between  log(Teff$_{HB}$) and
the  $\Gamma_{col}$  is very low and probably inside the errors 
(r$\simeq$0.14,  about  2\%   of the total   HB
morphology  variance).  This value increases up to r$\simeq$0.29, 
8\% of the total variance, if only intermediate metallicity clusters are considered.

The relation between log(Teff$_{HB}$) and $\Gamma_{col}$ is 
graphically presented in Figure
5, on the upper left panel. This result seems to suggest that
even if the  probability  of   stellar
collisions is higher, the HB morphology is not affected. 
Close encounters and  tidal stripping, suggested by
Fusi Pecci et  al.\, (1993) as a  possible origin  of bluer HB  stars,
do not seem to have a relevant role in HB morphology. 
Nevertheless, it is important to note that the collision rate
may have varied greatly through the clusters life time, especially in 
clusters which have undergone core collapse and re-expansion, and/or gravothermal
oscillations. Although none particular correlation for core-collapse
clusters in the sample has been noted, we should warn about the risk of
overinterpreting the lack of a good correlation between log(Teff$_{HB}$) 
and $\Gamma_{col}$.

\item \textbf{Central density}

The scenario is very similar to that inferred from the log(Teff$_{HB}$) -   
$\Gamma_{col}$ correlation. 
Figure 5, bottom left panel shows  the trend of HB temperature extension with
the central  density of  the cluster.  Our analysis indicates  a
weak correlation between both quantities: r$\simeq$0.16  for the complete clusters sample, and
r$\simeq$0.30, at intermediate metallicity [2.6\% and 9.0\% of
the total log(Teff$_{HB}$) variance respectively].  Again, contrary to previous studies (Fusi
Pecci et al.\, 1993), our sample indicates that cluster density is not
a ``good'' second  parameter, as it has  a little influence on HB
morphology.  Nevertheless, an equivalent caveat to that of  $\Gamma_{col}$ 
applies here. The central density now may not be as relevant as the 
maximum density achieved in the past.

\item \textbf{Other simple correlations}

Other quantities with which HB morphology seems to have a little but,
maybe still significant correlations are the distance to the Galactic
center, R$_{GC}$  (r$\simeq$0.19,  4\%  of the  total log(Teff$_{HB}$)
variance) and half-light  radius (r$\simeq$0.18, 3\%). The first  one,
may also be a secondary effect of the first  parameter, as $[Fe/H]$ has
an   already known trend   with  R$_{GC}$. 

\item \textbf{Age}

Finally, we have evaluated the influence of cluster age
in a subsample  of 47 clusters, in
common with the De Angeli et al.\ (2005)  
data base (see Table
1 and Figure 6).  The relative ages of the
47   clusters go  from 0.74 $\pm$ 0.06 for  the  youngest cluster  in the sample
(NGC~362) to 1.085 $\pm$ 0.06 for the  oldest one (NGC~7099).  These relative ages
are normalized to the average age  of the most metal poor clusters ($[Fe/H]  < -1.7$)
as explained by De Angeli et al.\    
The derived  Pearson  correlation  coefficient
between log(Teff$_{HB}$)  and age   is  r$=$ 0.04    (0.2\% of the    HB
morphology variance of   the clusters  subsample).
This result confirms the fact, already
pointed out by Rosenberg et  al.\ (1999), that age can  not be the only
explanation for the  second  parameter problem. 

In general, cluster age most important dependences are those with metallicity 
(r$=$ 0.38, 14\%) and Galactic latitude (r$=$ 0.42, 18\%).

However, it is interesting to point out that if we only consider intermediate
metallicity clusters (filled circles in Fig. 6), the correlation between log(Teff$_{HB}$) and cluster age
increases to r$=$0.76 (58\%). This will be explored later in more detail.

%\begin{table}[h]
%\begin{center}
%\begin{tabular}{ccc||ccc}
%\hline
%Cluster & log(Teff$_{HB}$) & $\bar{ \rm Age}$ & Cluster & log(Teff$_{HB}$) & $\bar{ \rm Age}$\\
%\hline
%\hline
%n0104  & 3.756  &  0.97 &  n6235 &  4.114  &  0.97 \\  
%n0362  & 4.079  &  0.74 &  n6266 &  4.477  &  0.92 \\  
%n1261  & 4.079  &  0.75 &  n6273 &  4.568  &  0.96 \\   
%n1851  & 4.097  &  0.80 &  n6284 &  4.279  &  0.86 \\
%n1904  & 4.352  &  0.90 &  n6287 &  4.114  &  1.05 \\
%n2808  & 4.568  &  0.77 &  n6342 &  3.778  &  0.92 \\ 
%n3201  & 4.079  &  0.77 &  n6356 &  3.756  &  0.97 \\  
%n4147  & 4.061  &  1.03 &  n6362 &  3.954  &  0.99 \\
%n4372  & 4.114  &  0.98 &  n6397 &  3.978  &  1.00 \\  
%n4590  & 4.041  &  0.94 &  n6544 &  4.176  &  0.84 \\
%n4833  & 4.301  &  1.01 &  n6584 &  4.041  &  0.85 \\
%n5024  & 4.079  &  1.02 &  n6624 &  3.771  &  0.88 \\
%n5634  & 4.146  &  0.98 &  n6637 &  3.748  &  0.89 \\
%n5694  & 4.204  &  1.05 &  n6638 &  4.097  &  0.87 \\
%n5824  & 4.380  &  1.02 &  n6652 &  4.000  &  0.89 \\
%n5904  & 4.176  &  0.96 &  n6681 &  4.301  &  1.01 \\
%n5927  & 3.724  &  0.94 &  n6717 &  4.114  &  0.92 \\
%n5946  & 4.279  &  0.90 &  n6723 &  4.130  &  1.00 \\
%n5986  & 4.415  &  0.91 &  n6838 &  3.763  &  0.84 \\
%n6093  & 4.477  &  1.04 &  n6864 &  4.176  &  0.85 \\  
%n6171 &  3.875  &  1.02 &  n6934 &  4.130  &  0.85 \\
%n6205 &  4.505  &  1.02 &  n6981 &  4.000  &  0.83 \\
%n6218 &  4.217  &  1.07 &  n7078 &  4.477  &  0.98 \\
%\hline
%\hline
%\end{tabular}
%\end{center}
%\caption{Relative ages from De Angeli et al.\ (2005) for the subsample of 47 clusters in common.}
%\end{table}

\end{itemize}

\section{The multivariate approach.}

While   the simple  approach    of examining  individual   monovariate
correlations  of log(Teff$_{HB}$) with cluster   parameters gives us a
good first look at the problem of HB morphology, the complexity of the
situation calls for a more sophisticated strategy. We are dealing with
a multidimensional data set, in which sets  of several observables may
be connected  in    multivariate  correlations. Simple,    monovariate
correlations  are only a  special  and rare  case. In particular, the
different trends  of log(Teff$_{HB}$) illustrated  above indicate that
the problem   of GC horizontal branch morphology is  intrinsically   statistically
multidimensional, and   that must  be  attacked  using  a multivariate
approach. 

% Ale: An~adir aqui un resumen sobre la tecnica de PCA.

First of   all, we  performed  the  Principal Component Analysis (PCA)  on  
our data set,   using all
independent input variables.  A code developed  at the Instituto de 
Astrof\' isica de Canarias (IAC) by A. Aparcio has been used.
We  ignored  the  derived  quantities,
$\rho_{0}$,     t$_{rc}$,  t$_{rh}$,  as  they   do    not  add to the
dimensionality of the data   manifold.    In addition, among  the  three
positional variables  R$_{GC}$,   {\it l}  and {\it  b},  we  have selected only
R$_{GC}$ and b. In  the same way, among the  group of variables formed
by $\Gamma_{col}$, $\mu_{V}$ and r$_{c}$, we took only $\Gamma_{col}$,
as  it  contains  the   other two  parameters   in   its formula.  
%The monovariate correlations for the 8 independent quantities explored are
%presented in Table 3. 
The input data were renormalized by subtracting the
mean and  dividing by the  sigma in each  of the  input variables.
The number of the significant eigenvalues, i.e.,
those  larger than expected   from  the measurement errors, gives  the
dimensionality of the data manifold.  Each of the eigenvectors also
accounts for a fraction of the  total sample variance. It is generally
agreed  that   eigenvalues $>$1  are    statistically significant,  but
somewhat lower ones  may be as well,  depending  on how the  data were
normalized.

%\begin{table}
%\begin{center}
%\footnotesize{
%\begin{tabular}{ccccccccc} 
%\hline
%\hline  
% & log(Teff$_{HB}$)& $[Fe/H]$ & M$_V$& $\Gamma_{col}$ & c & R$_{GC}$ & b & r$_h$\\
%\hline
%log(Teff$_{HB}$)&1.00&-0.54& -0.48 & 0.14&  0.09&  0.22&  0.09& -0.19  \\
%$[Fe/H]$  &    -0.54 & 1.00&  0.01 & 0.11& -0.10& -0.40& -0.21& -0.09 \\
%M$_V$  &       -0.48 & 0.01&  1.00 &-0.22& -0.08& -0.17&  0.06&  0.08 \\
%$\Gamma_{col}$& 0.14 & 0.11& -0.22 & 1.00&  0.66&  0.13& -0.28& -0.51 \\
%c   &           0.09 &-0.10&  0.08 & 0.66&  1.00&  0.06& -0.07& -0.27 \\
%R$_{GC}$ &      0.22 &-0.40& -0.18 & 0.13&  0.06&  1.00&  0.25& -0.32 \\
%B   &           0.09 &-0.21&  0.06 &-0.28& -0.07&  0.25&  1.00& -0.04 \\
%r$_h$  &       -0.19 &-0.09&  0.08 &-0.51& -0.27& -0.32& -0.04&  1.00\\
%\hline
%\end{tabular}}
%\end{center}
%\caption{Pearson  linear    regression  correlation
%coefficients, $\textbf{r}$, for the 8 independent quantities.}
%\end{table}

Table 4 presents the  eigenvalues (e$_i$),  fractional (V$_i$) and
cumulative  (C$_i$) contributions  to   the total sample  variance, in
percent,  for the obtained PCA  solution. Figure 7 shows the results
of  the PCA, as applied to  the entire above  set of independent input
variables,   in the form   of correlation-vector  diagrams.  
Usually, a steep drop  in the successive eigenvalues or in
the  fractional contributions to the   sample variance indicates where
the  number of statistically significant   dimensions stops, and where
the noise begins. Nevertheless, the situation is not always so clearcut.
In this data set, there could be at  least four, but probably as  
many as six statistically significant  dimensions or more.

The first four eigenvectors
account for the  78.8\% of the total sample  variance.  They  define a
natural coordinate system for this data set.  Projections of the input
axes  to the principal   planes   given by the eigenvectors   ($\xi_1,
\xi_2$) and   ($\xi_1,   \xi_3$) are    the
correlation-vector    diagrams   as shown   in  Figure 7. In this
representation, vectors  corresponding to  well-correlated   variables
define sharp or plane angles, on the contrary, uncorrelated quantities  
have orthogonal vectors.

 On the other hand, as in the previous Section, the situation for
intermediate metallicity clusters was also considered, by performing
PCA computations for the above described 8-parameters, including only
clusters with -1.8 $<$ [Fe/H] $<$ -1.3. In that case, the 100\% of the
data variability is explained with only 7 eigenvalues (the first four 
eigenvalues account for the 82.8\%), thus reflecting the
dimensionality decrease.

\begin{table}[h]
\begin{center}
\begin{tabular}{cccc} 
\hline
\hline  
  i & e$_i$ & V$_i$ & C$_i$ \\
\hline 
1& 2.32  &     29.0   &    29.0 \\
2& 1.85  &     23.1   &    52.1 \\
3& 1.18  &     14.8   &    66.9 \\
4& 0.95  &     11.9   &    78.8 \\
5& 0.71  &      8.9   &    87.7 \\
6& 0.61  &      7.6   &    95.3 \\
7& 0.22  &      2.7   &    98.0 \\
8& 0.16  &      2.0   &   100.0 \\
\hline 
\end{tabular}
\end{center}
\caption{Eigenvalues (e$_i$),  fractional (V$_i$) and
cumulative  (C$_i$) contributions  to   the total sample  variance, in
percent,  for the obtained PCA  solution for the set of 8 input independent
parameters.}
\end{table}

\begin{figure*}
\begin{center}
\begin{tabular}{c}
\includegraphics[width= 10 cm]{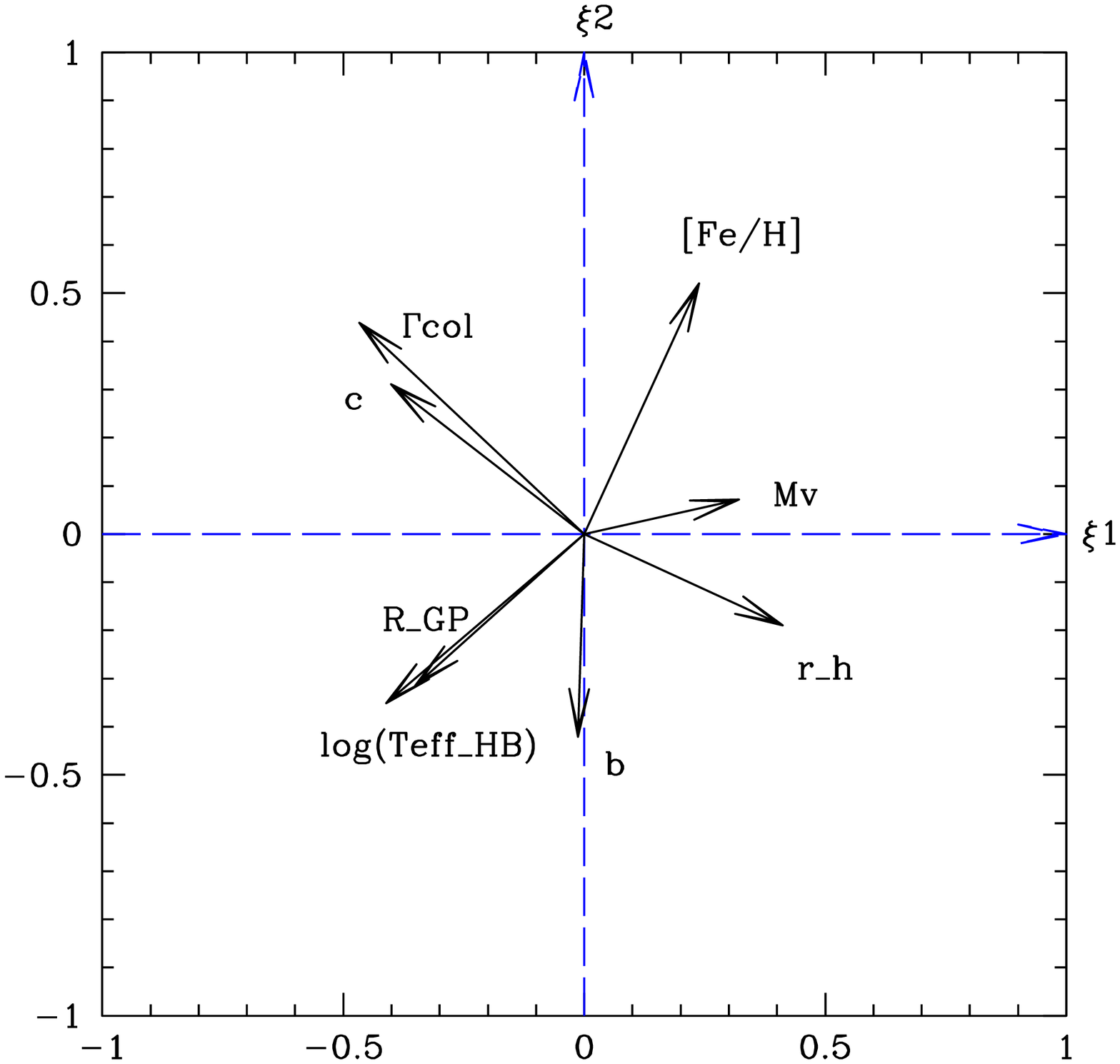}\\ 
\includegraphics[width= 10 cm]{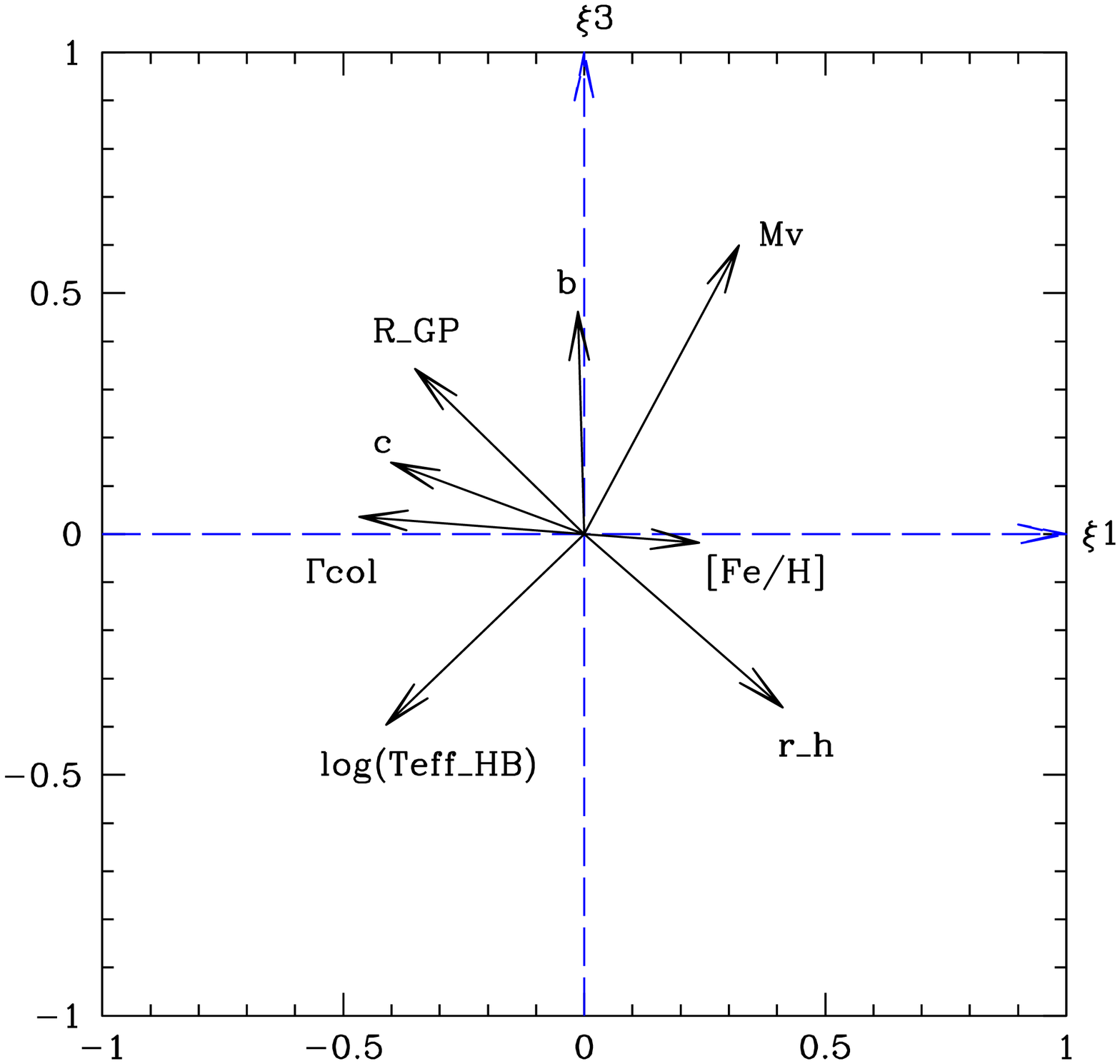}\\
\end{tabular}
\end{center}
\protect\caption[]
{Projections of the input
axes  to the principal   planes   given by the eigenvectors   ($\xi_1,
\xi_2$) and  ($\xi_1,   \xi_3$).}
%\label{firstp}
\end{figure*}

Obviously, the situation is quite  complex, reflecting the statistical
multidimensionality of the    entire  manifold  of   globular  cluster
properties (e.g. Djorgovski \&  Meylan, 1994). Given the  limited data set, a more
profitable approach is  to consider only   a subset of variables.   To
illustrate the point, we will consider only log(Teff$_{HB}$), $[Fe/H]$
and M$_V$. We find that for this data  subset of three input variables
(see Table 5),   at  least    two  statistical dimensions      are
necessary.  The first  two eigenvectors account  for  the 56\% and the
34\% of the total  sample  variance, respectively. The remaining  10\%
could  be accounted  by  the   errors, which   are quite difficult   to
evaluate as explained in  the previous sections.  If this is  true, a
weighted vector sum of $[Fe/H]$  and M$_V$ vectors could correlate  much
better with  log(Teff$_{HB}$)  than $[Fe/H]$ or  M$_{V}$ alone.  This is
shown in Figure 8 where the  bivariate correlation involving these 3
input quantities is:
 
\begin{center}

 $log$(Teff$_{HB}$) $= -0.17 \cdot $[Fe/H]$ - 0.136 \cdot $ M$_V + 2.84$

\end{center}

Clearly, the dispersion (rms$=$0.16) is now smaller than
the one observed in any of the monovariate correlations for the total
cluster sample. This bivariate correlation works specially well for
clusters with log(Teff$_{HB}$) $>$ 3.8 (Teff$_{HB}\geq$ 6000 K), that
is, clusters with at least some blue HB stars. In fact, if we perform
the previous bivariate analysis considering only these clusters, we
find a much clear correlation, with an rms$=$0.13, as shown in Figure 9
( we warn however of the smaller quantity of data in Fig~9).

\begin{figure*}
\centerline{\includegraphics[width= 11 cm]{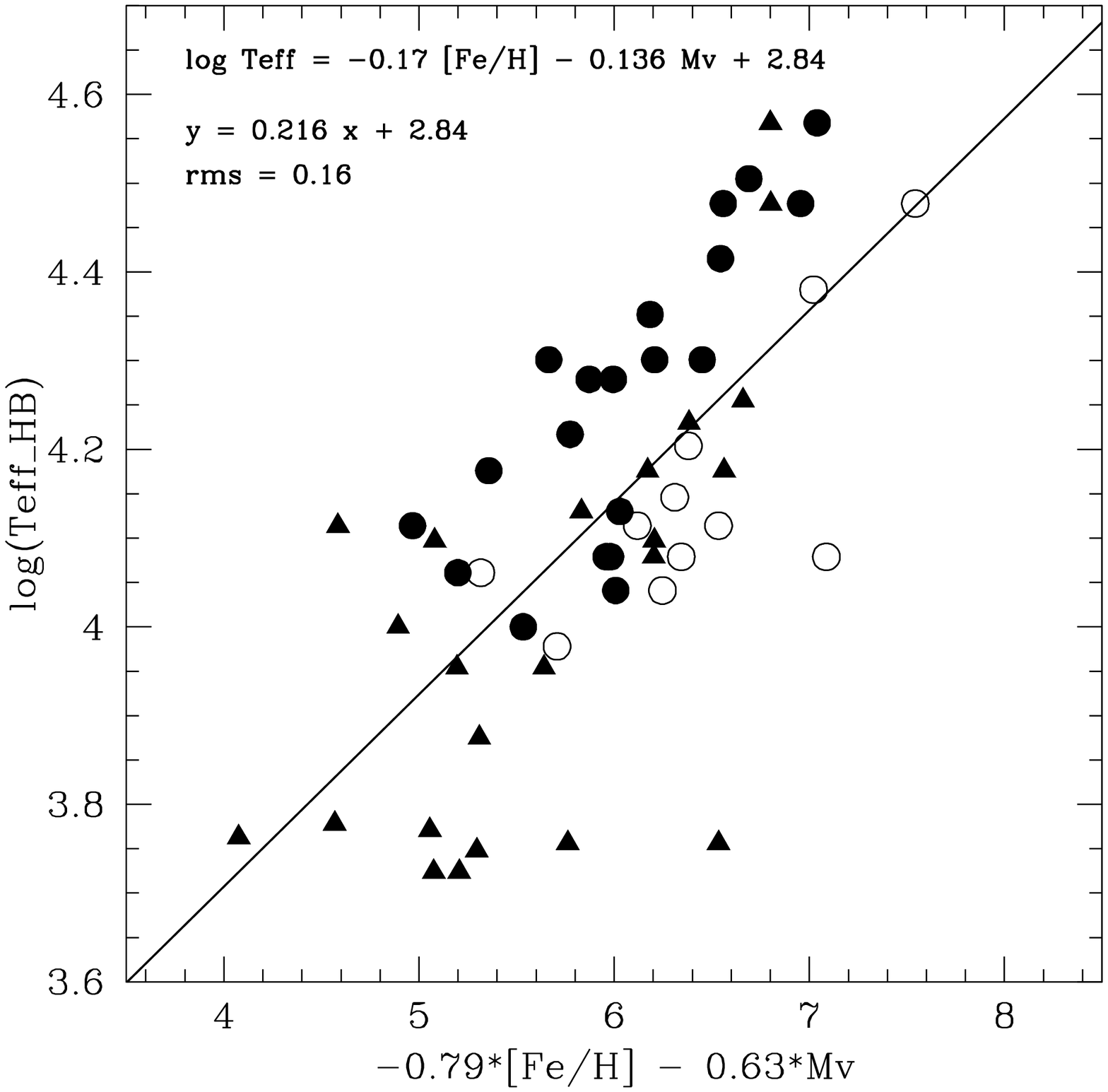}}
\protect\caption[]
{Bivariate   correlation   involving  the  variables  log(Teff$_{HB}$),
$[Fe/H]$ and M$_V$. The solid line is the least square fit to the data.  Symbols are those of Fig. 4}
\end{figure*}

\begin{table}
\begin{center}
\begin{tabular}{cccc} 
\hline
\hline  
  i & e$_i$ & V$_i$ & C$_i$ \\
\hline 
1  & 1.68     &   56.0    &    56.0\\
2  & 1.01     &   33.7    &    89.7\\
3  & 0.31     &   10.3    &   100.0\\
\hline 
\end{tabular}
\end{center}
\protect\caption[]
{PCA   solutions for a subsample  of  3  parameters : log(Teff$_{HB}$),
$[Fe/H]$ and M$_V$.}
\end{table}

\begin{figure*}
\centerline{\includegraphics[width= 11 cm]{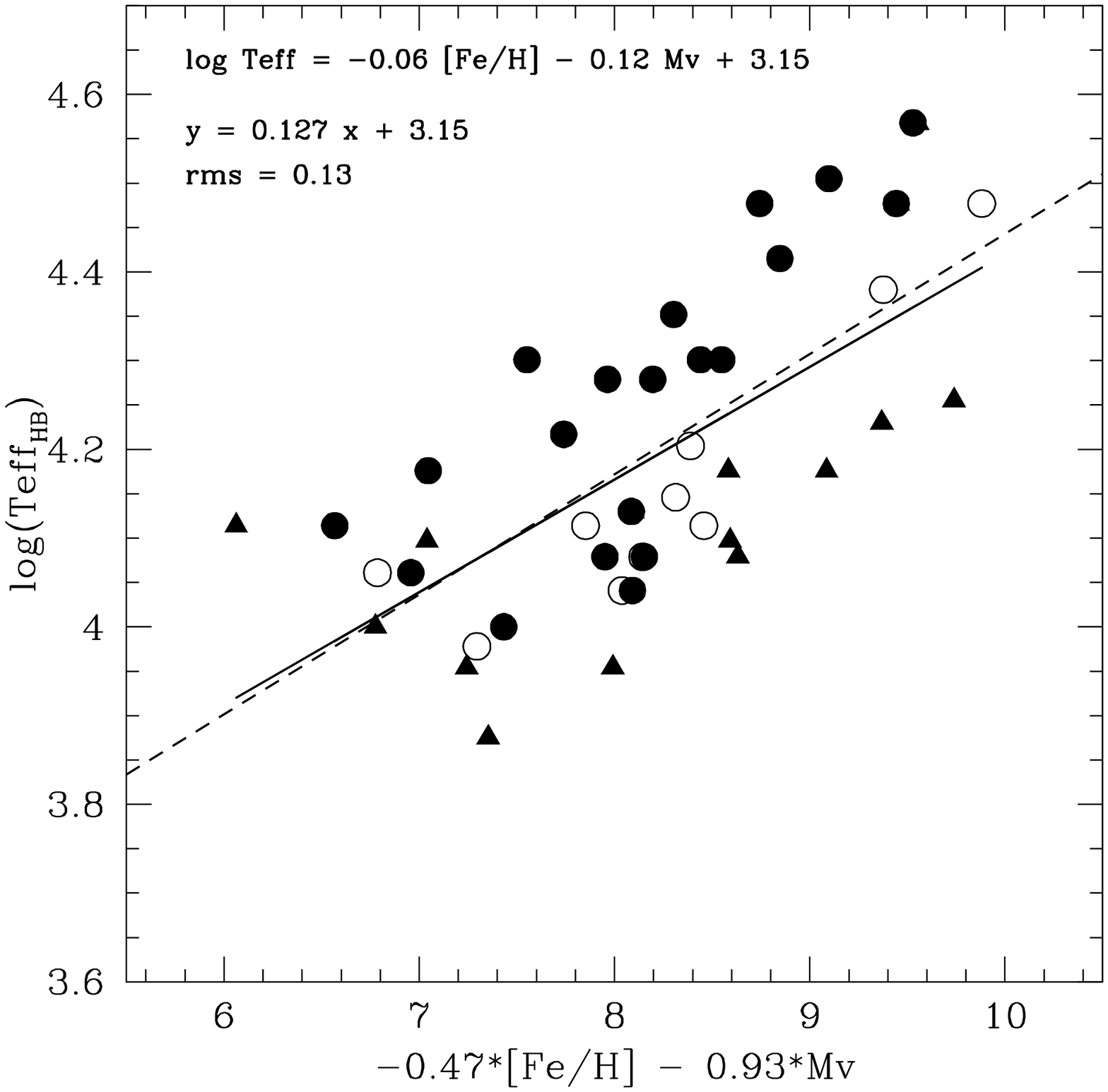}}
\protect\caption[]
{Bivariate correlation of  HB morphology with  optimized combinations
of $[Fe/H]$ and M$_V$ for clusters with log(Teff$_{HB}$) $>$ 3.8.
The plotted lines are the least square fit to the included data (solid)
and to clusters with $[Fe/H] \leq -1.3$ (dashed). Symbols are those of Fig. 4}.
\end{figure*}

On the other hand, we have also explored the trivariate correlation of
HB  morphology with  $[Fe/H]$, M$_V$ and  Age in order to possibly
reduce the   scatter  in  the  previous bivariate   relation,   due to 
possible effects of age. The analysis was performed
for a subsample of the 47 clusters in common with De Angeli et al.\ (2005)
which had  $[Fe/H] < -1.3$. 
The corresponding PCA results  are presented in Table
6 and Figure 10. The   third eigenvector  significance,   14.8\%,  increases with
respect to the combination of $[Fe/H]$ and M$_V$. Now, the first three
eigenvectors  account  for  the  93.0 \% of    the total variance. The
residual scatter is,   therefore,  of the  order  of the   measurement
errors. %After all, age might also influence the HB morphology.
%The spread of the clusters, NGC~104, NGC~362, NGC~1851, known
%to  be younger than the  mean (Rosenberg et al.\   1999), is, in fact,
%smaller than for the   bivariate relation (Figure 6).   The cluster
%NGC~7078 (plotted in red) is suspected to be related to a stream (Yoon
%\& Lee, 2002). On the other  hand, NGC~104 is  still very far from the
%general  trend of the  clusters, indicating that  some other mechanism
%has to be invocated, to explained its horizontal branch morphology.

\begin{table}
\begin{center}
\begin{tabular}{cccc} 
\hline
\hline  
  i & e$_i$ & V$_i$ & C$_i$ \\
\hline 
 1  &    1.86   &    46.5   &    46.5\\
 2  &    1.21   &    30.2   &    76.7\\
 3  &    0.69   &    17.3   &    94.0\\
 4  &    0.24   &     6.0   &   100.0\\
\hline 
\end{tabular}
\end{center}
\protect\caption[]
{PCA    solutions  for  the   subsample  of  4  parameters  :
log(Teff$_{HB}$), $[Fe/H]$, M$_V$ and Age.}
\end{table}

\begin{figure*}
\centerline{\includegraphics[width= 11 cm]{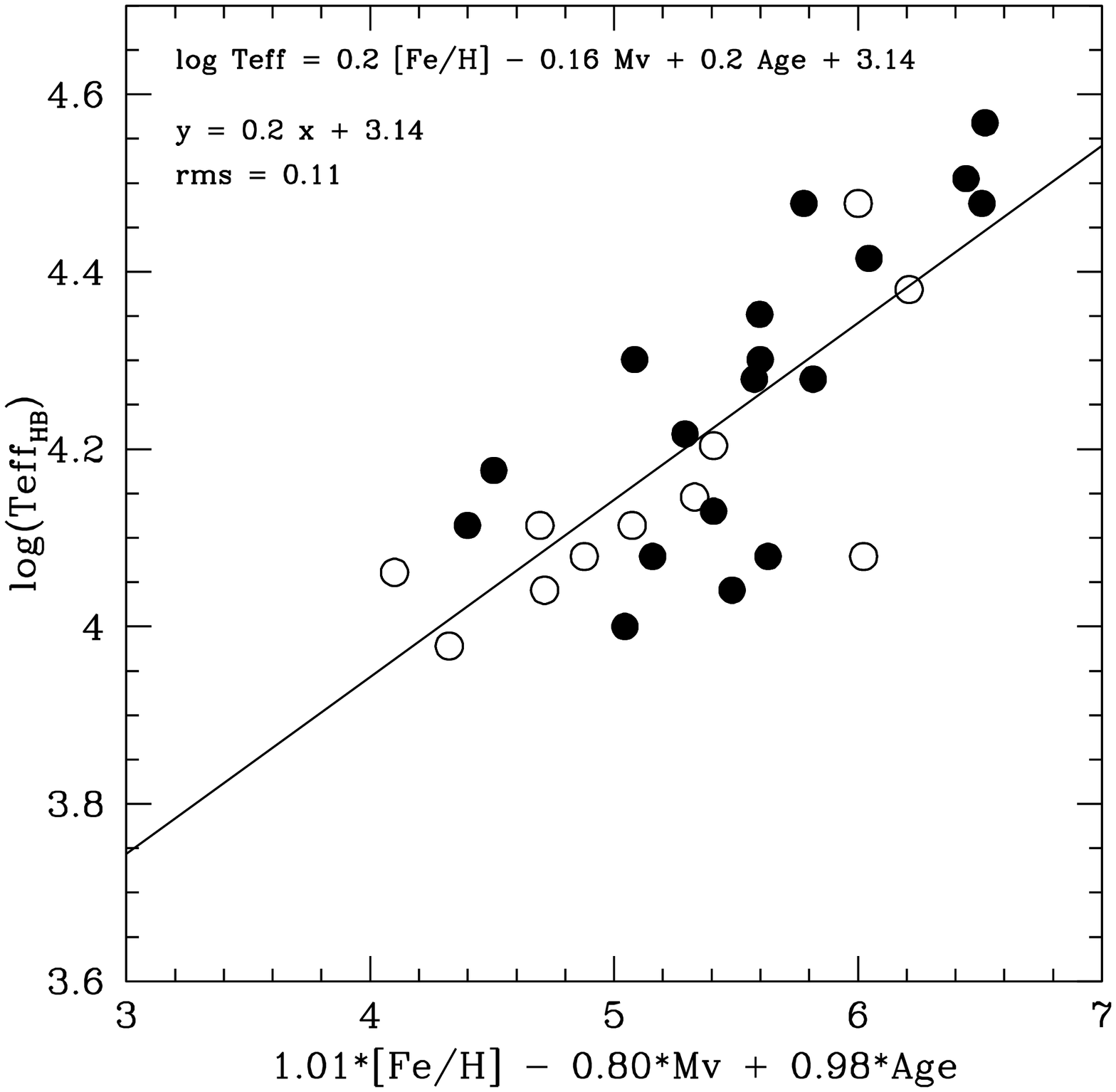}}
\protect\caption[]
{Trivariate correlations  of HB morphology with optimized combinations
of $[Fe/H]$, M$_V$  and Age for a  subsample of 28 clusters in common
with De Angeli et al.\ (2005) and $[Fe/H] \leq -1.3$. Symbols are those of Fig. 4}
\end{figure*}

\section{Summary and discussion: theoretical implications.}

In  this   section, the results   of  the
multivariate analysis presented above are summarized, trying to shed 
some light in their possible theoretical meaning.

As it has been extensively discussed by Djorgovski et al.\ (1993)  the GC manifold 
is rather complex, and this is perfectly true also when we 
focus our attention on the problem of the GC HB morphology and
of its dependence on the cluster parameters.
In the previous section, we have shown that at least four
parameters, but probably up to eight, are needed to
reproduce the HB extension in temperature. As Fusi Pecci, Buonnanno
and collaborators (see for example Fusi Pecci et al\., 1993), among others, 
have been remarking in the
last 20 years or so, there is no single ``second parameter'' which
can explain the HB anomalies, but a combination of parameters.

The present analysis, based on 54 GCs of the HST snapshot catalogue,
has given a few interesting new results which are worth of further
discussion.

At least   one new interesting  conclusion  can be  derived:  {\it the
importance of  total cluster luminosity, and  
therefore of total mass,
on  the  horizontal   branch   morphology}.   This effect, combined with  
the first parameter, can explain
probably  the major part of  the  Galactic globular cluster horizontal
branch morphologies. More   massive  clusters
(i.e.  more   luminous)  tend   to   have  more   extended  horizontal
branches. To  this scenario,  we have to add  the effect  of age, evaluated
here for 47  clusters in common with De Angeli et al.\ (2005).
 Lastly, the situation for intermediate-metallicity clusters
has also been analysed, leading to a higher correlation between the
temperature extension of the HB and M$_V$ (60\% of the total variation), 
with small increments of the $\Gamma_{col}$ and the central density 
contributions, that remain nevertheless inferior to 9\%.

One possible interpretation to the considerable influence of M$_V$ and
therefore, of cluster total mass, on HB morphology can be derived from
the paper by D'Antona et al.\  2002. 
They analyze the consequences,  on HB morphology, of helium  variation
due to self-pollution among globular cluster stars. Self-pollution had
already  been proposed      as an  explanation   for  the     chemical
inhomogeneities (spread in the   abundances of  CNO,  O  - Na and   Mg
anticorrelation) observed in GC members from the  main sequence to the
RGB  (see for  example Gratton  et al.\ 2001).  The ejecta  of massive
asymptotic giant  branch stars, which would  be the origin of the self
pollution,   would not    only  be  CNO processed,    but  also helium
enriched. D'Antona et al.\  (2002) models take 
into account this  possible helium enhancement with  respect to
the primordial value.   They find that a spread  in the helium content
does not affect the morphology of the  main sequence, turn off and RGB
in an easily  observable way. However,  the difference in the evolving
mass may play a role in the formation  of blue tails, as higher helium
stars  would be able  to populate much  bluer  HB regions. If  this is
correct,  self-pollution and  so helium  enrichment  would be higher in
more massive clusters,  as they would be able  to retain the  material
from the ejecta better than less massive  clusters. 

As already pointed out by Rosenberg, Recio-Blanco \& Garc\' ia-Mar\' in (2004),
M54, believed to be the remaining core of the Sagittarius dwarf galaxy
could be another example of this scenario.
More recently, the abundance analysis 
of stars in the double main sequence of  $\omega$ Centauri 
(Piotto et al., 2005) suggests
the presence of two populations of stars, one of which
is strongly He enhanced. This
could  be another observational indication supporting the D'Antona et al. 
theory in a very massive cluster.

Finally,  whatever is the  theoretical interpretation of the data,
a clear conclusion of this  analysis is that the influence of M$_V$ on
HB extension seems to be as important as those of metallicity and age.
Cluster total mass must be playing an important role in 
horizontal branch morphology.

\begin{acknowledgements}
A. Recio-Blanco thanks the support of the European Space Agency. GP and FDA acknowledge 
partial support by ASI and by MIUR under the program PRIN2003. This work was
supported in part by the STScI grants GO-6095, GO-7470, GO-8118, and GO-8723.

\end{acknowledgements}

\end{document}